\documentclass[onecolumn]{IEEEtran}
\usepackage{setspace,amsmath,latexsym,cite,amssymb,epsfig,amsfonts}
\usepackage{url,cite}
\usepackage{graphicx}
\usepackage{psfrag}
\usepackage{color}
\usepackage{epstopdf}
\usepackage{multirow}
\usepackage{standalone}
\usepackage{tikz}
\usepackage{pgfplots}
\pgfplotsset{compat=newest}
\allowdisplaybreaks

\title{Joint Antenna Selection and Spatial Switching for Energy Efficient MIMO SWIPT System}
\author{Jie~Tang,~\emph{Member},~\emph{IEEE}, Daniel~K.~C.~So,~\emph{Senior}~\emph{Member},~\emph{IEEE},\\Arman~Shojaeifard,~\emph{Member},~\emph{IEEE},~Kai-Kit Wong,~\emph{Fellow},~\emph{IEEE},\\ and Jinming Wen,~\emph{Member},~\emph{IEEE} 
\thanks{J. Tang is with the School of Electronic and Information Engineering, South China University of Technology, Guangzhou, China (eejtang@scut.edu.cn).\par D. K. C. So is with the School of Electrical and Electronic Engineering, University of Manchester, Manchester, United Kingdom (d.so@manchester.ac.uk). \par A. Shojaeifard and K.-K Wong are with the Department of Electronic and Electrical Engineering, University College London, London, United Kingdom (a.shojaeifard@ucl.ac.uk; kai-kit.wong@ucl.ac.uk). \par Jinming~Wen is with the Laboratoire de l'Informatique du Parall\'elisme, Universit\'e de Lyon, Lyon, France (e-mail: jinming.wen@ens-lyon.fr).}
\thanks{This work has been supported by the Xinghua Talents Program under grant J2RS-D615042III and the Engineering and Physical Sciences Research Council under grant EP/N008219/1.}
}

\begin{document}

\maketitle

\begin{abstract}

In this paper, we investigate joint antenna selection and spatial switching (SS) for quality-of-service (QoS)-constrained energy efficiency (EE) optimization in a multiple-input multiple-output (MIMO) simultaneous wireless information and power transfer (SWIPT) system. A practical linear power model taking into account the entire transmit-receive chain is accordingly utilized. The corresponding fractional-combinatorial and non-convex EE problem, involving joint optimization of eigen-channel assignment, power allocation, and active receive antenna set selection, subject to satisfying minimum sum-rate and power transfer constraints, is extremely difficult to solve directly. 
In order to tackle this, we separate the eigen-channel assignment and power allocation procedure with the antenna selection functionality. In particular, we first tackle the EE maximization problem under fixed receive antenna set using Dinkelbach-based convex programming, iterative joint eigen-channel assignment and power allocation, and low-complexity multi-objective optimization (MOO)-based approach. On the other hand, the number of active receive antennas induces a trade-off in the achievable sum-rate and power transfer versus the transmit-independent power consumption. We provide a fundamental study of the achievable EE with antenna selection and accordingly develop dynamic optimal exhaustive search and Frobenius-norm-based schemes. Simulation results confirm the theoretical findings and demonstrate that the proposed resource allocation algorithms can efficiently approach the optimal EE.
\end{abstract}


\section{Introduction}

Ambient radio frequency (RF) signals can be used in conjuction with transmitting information to transfer power with adequate efficiency over relatively short transmit-receive distances \cite{6623062}. As a result, wireless power transfer (WPT) and energy harvesting (EH) have recently emerged as promising candidate solutions for jointly improving energy efficiency (EE) and prolonging battery-life in fifth-generation (5G) and beyond communication systems \cite{6951347}. Simultaneous wireless information and power transfer (SWIPT) is considered particularly attractive for small-cell networks and device-to-device (D2D) communications \cite{6845056}.

The information theoretic bounds for a single-input single-output (SISO) SWIPT system was investigated in \cite{4595260}. In particular, a capacity-energy function was developed under the assumption that the receiver can simultaneously perform information decoding (ID) and EH from the same RF signal without any limitations. Despite being insightful, such theoretical bounds are not practically feasible considering that the information and energy receivers sensitivities are fundamentally different under current technologies \cite{6845056}. Motivated by this, the authors in \cite{6489506} investigated practical beamforming techniques in a multiple-input multiple-output (MIMO) SWIPT system and proposed two potential receiver design strategies, namely time-switching (TS) and power-splitting (PS). Based on these, many recent research works have been carried out considering different system aspects. 

In \cite{Xiangz2012}, the authors proposed a worst-case robust beamforming design with a virtual TS-based receiver. In \cite{Liu2013}, the authors further derived the optimal mode switching rule at the receiver based on the TS technique. In particular, the trade-off between information decoding (ID) and EH, characterized as the boundary of a so-called `outage energy' region, was exploited in \cite{Liu2013}. On the other hand, the authors in \cite{AANasir} proposed a PS-based relaying protocol in order to maximize the throughput whilst performing ID and EH at the relay. The work in \cite{Liu2013c} studied a point-to-point wireless link over flat-fading Rayleigh channel, where the receiver was assumed to primarily rely on EH from the transmitted RF signals in order to function. Accordingly, an approach for jointly optimizing transmit power and receiver PS using the trade-off between ID and EH was provided in \cite{Liu2013c}. In addition to these results, SWIPT has been studied in the context of multi-user systems in \cite{ZhouX14}, considering both time division multiple access (TDMA) and orthogonal frequency division multiple access (OFDMA). In particular, for the TDMA-based information transmission, the TS technique was applied at the receivers, whereas PS was employed for the receivers in the OFDMA-based counterpart. In contrast to the conventional TS and PS approaches, a new technique called spatial switching (SS) for a point-to-point MIMO SWIPT system with RF EH capabilities was recently proposed in \cite{Timotheou2015}. The proposed technique utilizes the spatial MIMO channel structure through singular value decomposition (SVD) with eigen-channels representing either the transport of information or the transfer of energy. 

The focus in most existing works on SWIPT systems has been placed on maximizing either the throughput or the harvested energy. However, designing systems with the sole goal of spectral efficiency (SE) maximization constitutes to ever-rising network power consumption, which goes against global commitments for sustainable development \cite{TangJSAC}. Meanwhile, the alternative approach to harvest as much energy as possible adversely affects information transfer, leading to the degradation of system quality of service (QoS). On the other hand, EE, is considered an increasingly important metric in the design of energy efficient communication networks \cite{Hasan2011}. In fact, many works on green cellular network design have emerged in recent years, see, e.g., \cite{Chen2011,Miao2010,TangRE,6918448,TangTCOM, armanEE}. Furthermore, the EE maximization is currently an active research topic for SWIPT systems \cite{QShi, HeS,6774838,7322196,TangsubmitTVT}. The state-of-the-art works on the topic so far are however mostly based on either PS or TS receiver techniques. Hence, in this paper, we provide a fundamental study of the EE optimization problem considering a SS-based MIMO SWIPT system.

\subsection{Contributions}

The original work on SS-based SWIPT in \cite{Timotheou2015} aims to minimize the transmit power, this approach however may not be energy efficient considering the situation where the overall power consumption is dominated by the circuit power consumption \cite{Chen2011, Xiong2011}. Therefore, in this paper, we investigate the EE maximization problem in a SS-based MIMO SWIPT system with a practical power model where the number of active receive antennas, transmit power as well as harvested energy are taken into consideration. Our aim is to maximize the EE under minimum sum-rate and power transfer constraints, by jointly optimizing receive active antenna set selection, eigen-channel assignment, and power allocation. The EE optimization problem under consideration is extremely difficult to tackle directly, given that it is fractional-combinatorial and non-convex. In order to tackle this problem, we propose a dual-layer approach where the antenna selection procedure and the eigen-channel allocation and power allocation operation are separated. For a fixed receive antenna set (inner-layer), a near-optimal upper-bound resource allocation approach based on the Dinkelbach method is developed. In particular, we need to update $\beta$ through the Dinkelbach method and apply convex programming-based solution for each iteration. In addition, we propose a novel low-complexity iterative resource allocation approach in order to jointly assign eigen-channels and allocate transmit power. To further reduce the computational complexity, we propose a heuristic algorithm based on the idea of multi-objective optimization. Meanwhile, in order to further explore the EE of the SS-based MIMO SWIPT system, the receive antenna selection strategy (outer-layer) is fundamentally investigated. Intuitively, activating more receive antennas allows for achieving higher sum-rate and harvest energy, this however comes at the cost of larger transmit-independent power consumption. Hence, based on the idea of multi-objective optimization, we propose a Frobenius-norm-based antenna selection scheme to exploit the trade-off between the sum-rate gain and the overall power consumption towards improving the achievable EE. We provide simulation results in order to confirm the validity of our theoretical findings and draw design insights into the performance of SS-based MIMO SWIPT systems. 

\subsection{Organization and Notation}

The remainder of this paper is organized as follows. The system model and problem formulation is given in Section II. In Section III, joint eigen-channel assignment and power allocation schemes for the SS-based MIMO SWIPT system are proposed under fixed receive antenna set. In Section IV, we further investigate the achievable EE using antenna selection and develop optimal exhaustive search and Frobenius-norm-based schemes. Simulation results are provided in Section V and conclusions are drawn in Section VI.

The following notations are used throughout the paper. Bold upper and lower case letters respectively represent matrices and vectors; $(\cdot)^H$ denotes the matrix conjugate transpose; $\textbf{I}_{N_t \times N_t}$ corresponds to an $N_t \times N_t$ identity matrix; $\textmd{Tr}(\cdot)$ is the trace of a matrix; and $[x]^{+}$ stands for $\max(x, 0)$.

\section{Preliminaries}

In this section, we first describe the MIMO SWIPT system model with SS-based receiver and then mathematically formulate the EE optimization problem.

\subsection{System Model}
\begin{figure}\centering
 \includegraphics[width=0.8\columnwidth]{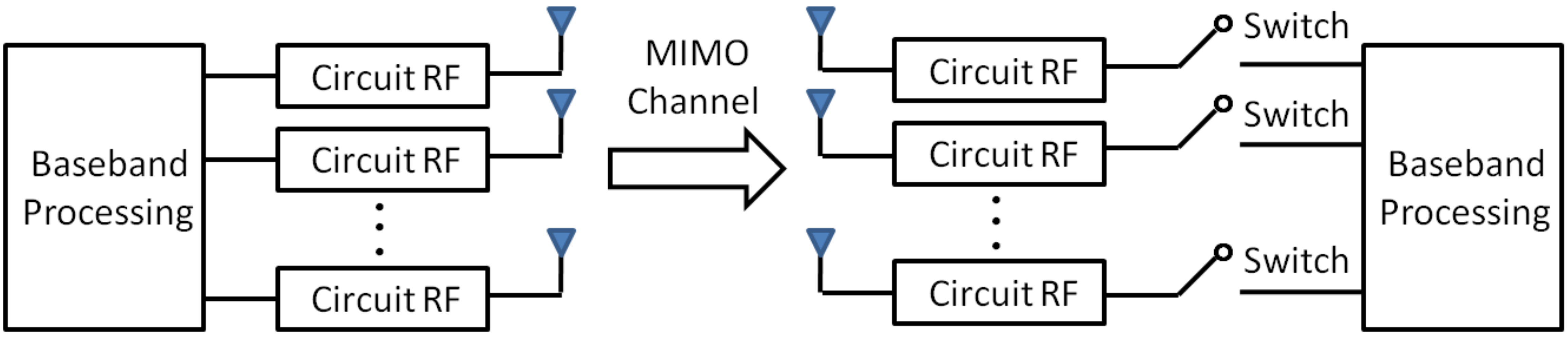}
 \caption{Schematic example of a point-to-point MIMO system with SS-based receiver.} 
 \label{system_model}
\end{figure}

We consider a point-to-point MIMO SWIPT system where the source and the destination are respectively equipped with $N_T$ transmit antennas and $N_R$ receive antennas, as shown in Fig. \ref{system_model}. We assume a constant power supply is connected to the source, whilst the destination is capable of harvesting and transferring RF energy. In the context of SS-based receiver, the MIMO channel can be decomposed using SVD with the corresponding eigen-channels being used either to convey information or to transfer energy \cite{Timotheou2015}. 

Intuitively, employing more receive antennas allows for achieving higher sum-rate and harvested energy. This however comes at the cost of larger transmit-independent power consumption. As a result of this trade-off, fully utilizing all available receive antennas with SS receiver does not necessarily correspond to an energy efficient strategy. In fact, the appropriate selection of the active receive antenna set through activation/deactivation of the corresponding RF chain switches is essential towards achieving high EE. With all receive antennas active, the channel matrix from the source to destination is denoted with $\textbf{H} \in C^{N_R \times N_T}$. In this work, we consider an uncorrelated flat-fading MIMO Rayleigh channel model. As a result, with the number of active receive antennas $N$, the selected active receive antenna set and the corresponding channel from the source to the destination are respectively denoted with $\chi \in \{1,\cdots,N_R\}$ and $\textbf{H}_\chi \in C^{N \times N_T}$, where $N = |\chi|$. Let $\textbf{x} \in C^{N_T \times 1}$ and $\textbf{n} \in C^{N \times 1}$ denote the transmit signal vector and circularly symmetric complex additive white Gaussian noise (AWGN) vector with zero mean and unit variance, respectively.

The received signal can be expressed as
\begin{equation}\label{received signal}
    {\textbf{y} = \textbf{H}_{\chi} \textbf{x}+ \textbf{n},}
\end{equation}
where $E[\textbf{x}\textbf{x}^H] = \textbf{Q}_\chi$, with $\textbf{Q}_\chi$ being the transmit covariance matrix \cite{7478073}. Therefore, with the selected receive antenna set $\chi$ ($\textbf{H}_\chi \in C^{N \times N_T}$), the mutual information (MI) in the MIMO SWIPT system with SS-based receiver is formulated as \cite{Goldsmith2005}
\begin{equation}\label{rate}
{I(\textbf{x};\textbf{y}) = \log\det(\textbf{I}_N + \textbf{H}_\chi \textbf{Q}_\chi \textbf{H}_\chi^H).}
\end{equation}

\begin{figure}\centering
 \includegraphics[width=0.8\columnwidth]{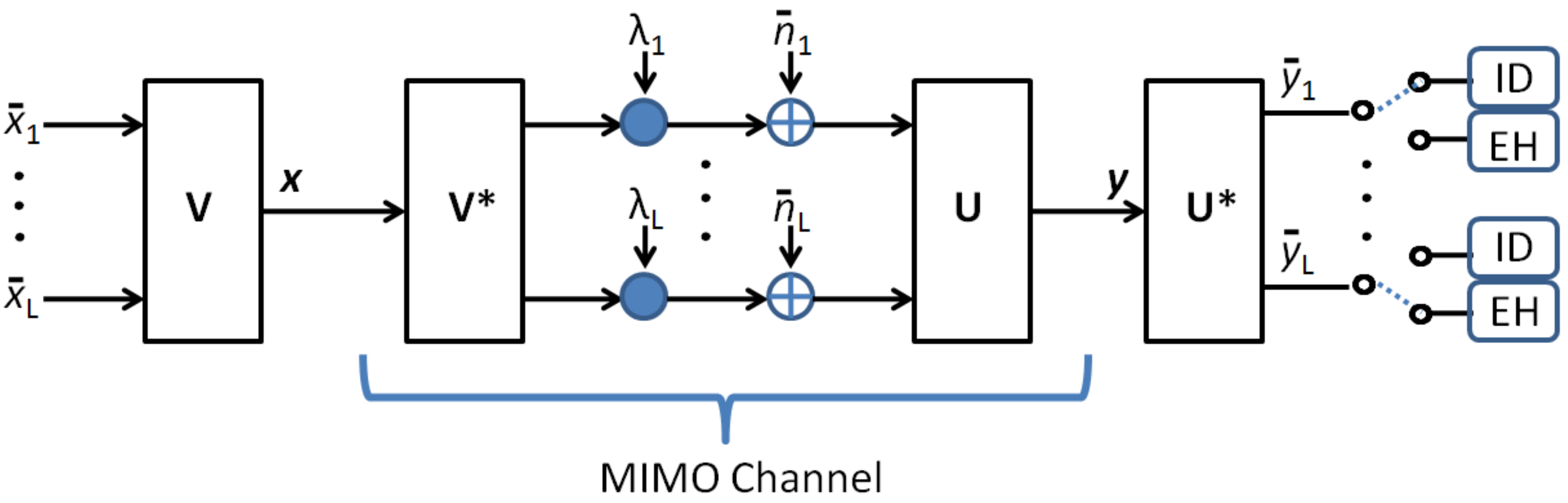}
 \caption{Schematic example of the SVD of the MIMO channel into $L$ parallel AWGN channels.} 
 \label{SVD}
\end{figure}

Fig. \ref{SVD} provides an illustrative example of the MIMO channel decomposition for potentially conveying information and energy. With a fixed active receive antenna set $\chi$, the SVD-based transformation of the channel matrix $\textbf{H}_\chi$ can be expressed as $\textbf{H}_\chi = \textbf{U} \mathbf{\Sigma} \textbf{V}^H$, where $\textbf{U} \in C^{N \times N}$ and $\textbf{V} \in C^{N_T \times N_T}$ correspond to unitary matrices whilst $\mathbf{\Sigma} \in C^{N \times N_T}$ is a diagonal matrix containing the singular values of the channel matrix $\textbf{H}_\chi$, $\lambda_i (\chi)$, respectively. Hence, the MIMO channel (with the selected receive antenna set $\chi$) is decomposed into $L$ parallel SISO channels with \cite{7277029}
\begin{equation}\label{received signal}
    {\tilde{y}_i =\lambda_i (\chi) {\tilde{x}}_i+ {\tilde{n}}_i,}
\end{equation}
where ${\tilde{n}}_i$ is AWGN for the \textit{i}-th parallel SISO channel. Considering that SVD is a unitary transformation of the MIMO channel, ${\tilde{n}}_i$ follows the same distribution with that of ${n}_i$. Therefore, as illustrated in Fig. \ref{SVD}, the output of each eigen-channel is connected either to the ID circuit or to the EH rectification circuit.

In order to depict the above notion, we use the binary variable $\alpha_i$ in order to indicate whether the \textit{i}-th eigen-channel is used for data transmission ($\alpha_i = 1$) or energy transfer ($\alpha_i = 0$). Therefore, the sum-rate of the $L$ parallel SISO channels with under the selected receive antenna set $\chi$ is given by 
\begin{equation}\label{rate}
{C = \sum_{i=1}^{L} \log_2 (1+ \alpha_i p_i \lambda_i (\chi))}
\end{equation}
where $p_i$ is the power allocated to the \textit{i}-th eigen-channel for data transmission. On the other hand, the total harvested energy at the receiver can be written as
\begin{equation}\label{harvesting energy}
{E= \sum_{i=1}^{L} \eta (1-\alpha_i) p_i \lambda_i (\chi)}
\end{equation}
where $\eta$ is a constant representing the loss from the energy transducer conversion of the harvested energy to electrical energy.

\subsection{Power Model}

For the SS-based MIMO SWIPT system under consideration, the power model should account for the power consumption of the entire transmit-receiver chain. This includes the impact of the transmit power, circuit power, as well as RF energy harvester. It can be argued that the latter consumes small amounts of power, and thus may not significantly affect the EE of the system. On the other hand, it is intuitive to infer that the system power consumption is in part may be compensated by the transferred energy. As a result, similar to the approach in \cite{6781609}, here, the harvested energy is taken into consideration. In particular, the total power consumption is formulated using a linear power model as follows
\begin{equation}\label{overall power}
{P = \zeta P_T + P_C - E}
\end{equation}
where $\zeta$, $P_T$ and $P_C$ are respectively represent the reciprocal of drain efficiency of the power amplifier, transmit power, and the total circuit power consumption (note that $E$ is defined in (\ref{harvesting energy})). The minus sign in (\ref{overall power}) implies that the receiver is able to harvest a portion of the radiated power from the transmitter. The total transmission power $P_T$ correspond to the sum of all powers allocated to the eigen-channels, i.e. $P_T = \sum_{i=1}^{L}p_i$. Further, the total circuit power consumption $P_C$ can be split into static and dynamic parts based on the configurations of the active links. In this work, the transmit-dependent circuit power consumption is modeled as a linear function of the number of active antennas using
\begin{equation}\label{circuit_power}
{ P_C = P_{sta} + P_{ant}^{BS} N_T + P_{ant}N = \bar{P}_{sta} + P_{ant}N }
\end{equation}
where $\bar{P}_{sta} = P_{sta} + P_{ant}^{BS} N_T$ is the static circuit power at the transmitter and $P_{ant}N$ denotes the dynamic power consumption which is proportional to the number of active receive antennas in a SS-based MIMO SWIPT system. In addition, although transmit antenna selection can in theory improve the EE performance in a conventional MIMO system \cite{TangTVT}, SS for deciding on the ID and EH is carried out at the receiver side. We therefore only consider the antenna selection operation at the receiver side in this work.

\subsection{EE Optimization Problem}

The EE can be defined as the total number of delivered bits per unit energy. Hence, we express the EE of the SS-based MIMO SWIPT system with receive antenna set $\chi$ using
\begin{equation}\label{EE}
{\psi_{EE} \triangleq \frac{C}{P} = \frac{\sum_{i=1}^{L} \log_2 (1+ \alpha_i p_i \lambda_i (\chi))}{\zeta P_T + \bar{P}_{sta} + P_{ant}N - \sum_{i=1}^{L} \eta (1-\alpha_i) p_i \lambda_i (\chi)},}
\end{equation}
where $C$ is the corresponding sum-rate of the $L$ parallel SISO channels with the selected receive antenna set $\chi$. We can now proceed to optimization problem formulation.

The objective of this paper is to maximize the EE of a SS-based MIMO SWIPT system whilst meeting two important QoS constraints in terms of minimum sum-rate and harvested energy. The corresponding optimization problem can be mathematically formulated as
\begin{eqnarray}
\label{objective} && \max_{\alpha_i,p_i,\chi}~ \frac{\sum_{i=1}^{L} \log_2 (1+ \alpha_i p_i \lambda_i (\chi))}{\zeta \sum_{i=1}^{L}p_i + \bar{P}_{sta} + P_{ant}N - \sum_{i=1}^{L} \eta (1-\alpha_i) p_i \lambda_i (\chi)}\\
\label{constraint1} && \textrm{s.t.} ~~ \sum_{i=1}^{L} \log_2 (1+ \alpha_i p_i \lambda_i (\chi)) \geq R_{min},\\
\label{constraint2} && ~~~~~ \sum_{i=1}^{L} \eta (1-\alpha_i) p_i \lambda_i (\chi)) \geq E_{min},\\
\label{constraint3} &&~~~~~ \sum_{i=1}^{L}p_i \leq P_{{max}},\\
\label{constraint4} &&~~~~~~ p_i \geq 0, \alpha_i \in \{0,1\}, \forall i \in \mathcal{L},
\end{eqnarray}
where $R_{min}$, $E_{min}$ and $P_{{max}}$ are respectively the minimum sum-rate, minimum harvested energy, and maximum transmit power constraints. It should be noted that the analysis is not restricted based on energy storage availability. The minimum energy requirement is defined as the minimum additional amount of harvested energy in one transmission cycle in cases where energy storage is not viable. Otherwise, $E_{min}$ corresponds to the required amount of energy of the energy harvester to function in the next transmission cycle. Moreover, (\ref{constraint3}) and (\ref{constraint4}) are the constraints for the allocated power and SS indicators. 

It is easy to see that the EE optimization problem involves binary and continuous variables as well as non-linear functions; hence it belongs to the class of mixed-integer non-linear optimization problems. Furthermore, jointly selecting the ``best'' receive antenna set $\chi$, eigen-channel assignment $\alpha_i$, and power allocation $p_i$ makes the problem (\ref{objective})-(\ref{constraint4}) non-convex and hence intractable to tackle directly. Consequently, in the following sections, we develop joint antenna selection and SS approaches in order to maximize EE. Since $\chi$ affects the EE optimization problem in a comprehensive manner, i.e., $\chi$ relates to the channel matrix (effective channel gain), the dynamic spatial assignment and power allocation, solving $\chi$ jointly with $\alpha_i$ and $p_i$ is not straightforward. Nevertheless, for any optimization problems, it is possible to tackle the problem over some of the variables and then over the remaining ones \cite{Boyd04}. Therefore, we will optimize the eigen-channel assignment $\alpha_i$ and power allocation $p_i$ at first (inner-layer process) under fixed receive antenna set $\chi$. Thereafter, we propose a strategy to determine the optimal receive antenna set $\chi$ in order to further improve the achievable EE (outer-layer process).

\section{Joint Eigen-Channel Assignment and Power Allocation}

In this section, we consider joint eigen-channel assignment and power allocation algorithm for the SS-based MIMO SWIPT system under fixed antenna set $\chi$. In particular, we develop three approaches for the inner-layer process. First, we develop a near-optimal solution based on the Dinkelbach method and convex programming. Since the complexity of convex programming is comparatively high, we propose a novel iterative resource allocation approach where the eigen-channel assignment functionality and power allocation operation are separated. To further reduce the computational complexity, a sub-optimal solution based on multi-objective optimization is proposed. For ease of description, we omit $\chi$ in the subscript in this section. Even though the antenna set is fixed here, the problem still belongs to the class of mixed-integer non-linear optimization problems, which is very difficult to solve directly. Similar to the approximation widely used in the context of OFDMA resource allocation \cite{Xiong2011}, our eigen-channel assignment and power allocation problem can be approximated as
\begin{eqnarray}
\label{objectivefa} && \max_{\tilde{\alpha}_i,p_i}~ \frac{\sum_{i=1}^{L} \tilde{\alpha}_i \log_2 (1+ \frac{p_i \lambda_i}{\tilde{\alpha}_i} )}{\zeta \sum_{i=1}^{L}p_i + \bar{P}_{sta} + P_{ant}N - \sum_{i=1}^{L} \eta (1-\tilde{\alpha}_i) p_i \lambda_i}\\
\label{constraint1fa} && \textrm{s.t.} ~~ \sum_{i=1}^{L} \tilde{\alpha}_i \log_2 \left( 1+ \frac{p_i \tilde{\lambda}_i}{\tilde{\alpha}_i} \right) \geq R_{min},\\
\label{constraint2fa} && ~~~~~ \sum_{i=1}^{L} \eta (1-\tilde{\alpha}_i) p_i \lambda_i  \geq E_{min},\\
\label{constraint3fa} &&~~~~~ \sum_{i=1}^{L}p_i \leq P_{{max}},\\
\label{constraint4fa} &&~~~~~~ p_i \geq 0, \tilde{\alpha}_i \in [0,1], \forall i \in \mathcal{L}.
\end{eqnarray}
It should be noted that when $\tilde{\alpha}_i$ approaches zero, $\tilde{\alpha}_i \log_2 (1+ \frac{p_i \tilde{\lambda}_i}{\tilde{\alpha}_i} )$  also tends to zero, which is similar to setting $\alpha_i$ to zero , i.e., the \textit{i}-th eigen-channel is nearly not assigned for data transmission but for EH. On the other hand, when $\tilde{\alpha}_i$ is close to one, $\tilde{\alpha}_i \log_2 (1+ \frac{p_i \tilde{\lambda}_i}{\tilde{\alpha}_i} )$ is close to $\log_2 (1+ \alpha_i p_i \lambda_i )$, which indicates that the \textit{i}-th eigen-channel is almost entirely assigned for data transmission. Therefore, when $\tilde{\alpha}_i$ is close to zero or one, the approximation becomes precise. As a result, we will use $\tilde{\alpha}_i$ instead of $\alpha_i$ to represent the eigen-channel assignment for either data transmission or EH of the \textit{i}-th channel in a modified EE optimization problem. On the other hand, it should also be noted that the solution of problem (\ref{objectivefa})-(\ref{constraint4fa}) may provide fractional eigen-channel assignment $\tilde{\alpha}_i^*$, and hence the proposed transformation provides an upper-bound solution.

It should also be noted that the optimal solution involves eigen-channel assignment $\tilde{\alpha}_i^*$ are not strictly either 0 or 1. To get a feasible solution to the original optimization problem, we need to round the possibly fractional eigen-channel assignment $\tilde{\alpha}_i^*$ to 0 or 1 and then perform the power allocation algorithm to get the maximum ``reasonable'' EE for the round-off $\tilde{\alpha}_i^{round}$. On the other hand, it has been shown in \cite{Yu2002} that the optimal $\tilde{\alpha}_i^*$ mostly tends to either 0 or 1, hence this enables us to precisely solve the original problem.

\subsection{Convex Programming-based Dinkelbach Method (DM-CVX)}

Since the optimization problem in (\ref{objectivefa})-(\ref{constraint4fa}) involves a non-linear fractional programming problem, it is non-convex and difficult to solve directly. However, given that the Dinkelbach method is an efficient method to tackle such problems \cite{Dinkelbach1967}, we therefore can apply it to solve our non-convex non-linear fractional programming problem. Specifically, we transform the fractional form objective function into a numerator-denominator subtractive form using the following proposition.

\textbf{\emph{Proposition 1:}} \emph{The maximum achievable EE $\beta^* = \psi_{EE}^*$ can be obtained provided that}
\begin{equation}
\label{Dinkelbach2} {\max_{\textbf{p}, \tilde{\boldsymbol{\alpha}}}~ U_R(\textbf{p}, \tilde{\boldsymbol{\alpha}}) - \beta^* U_T(\textbf{p}, \tilde{\boldsymbol{\alpha}}) =  U_R(\textbf{p}^*, \tilde{\boldsymbol{\alpha}}^*) - \beta^* U_T(\textbf{p}^*, \tilde{\boldsymbol{\alpha}}^*) = 0}
\end{equation}
for $U_R(\textbf{p}, \tilde{\boldsymbol{\alpha}}) \geq 0$ and $U_T(\textbf{p}, \tilde{\boldsymbol{\alpha}}) \geq 0$, where
\begin{eqnarray}
\label{Dinkelbach3} && U_R(\textbf{p}, \tilde{\boldsymbol{\alpha}}) = \sum_{i=1}^{L} \log_2 (1+ \alpha_i p_i \lambda_i ), \\
\label{Dinkelbach4} && U_T(\textbf{p}, \tilde{\boldsymbol{\alpha}}) = \zeta \sum_{i=1}^{L}p_i + P_{sta} + P_{ant}N - \sum_{i=1}^{L} \eta (1-\alpha_i) p_i \lambda_i,\\
\label{Dinkelbach55} && \beta^* = \frac{U_R(\textbf{p}^*, \tilde{\boldsymbol{\alpha}}^*)}{U_T(\textbf{p}^*, \tilde{\boldsymbol{\alpha}}^*)}.\\
\label{Dinkelbach5} && \textrm{and} ~~ \textbf{p} = [p_0, p_1,\cdots, p_L], \tilde{\boldsymbol{\alpha}} = [\tilde{\alpha}_1, \tilde{\alpha}_2,\cdots,\tilde{\alpha}_L].
\end{eqnarray}
\emph{Proof:} Please refer to \cite{Dinkelbach1967} for a proof of \emph{Proposition 1}.

\emph{Proposition 1} provides an adequate and compulsory condition for developing the optimal resource allocation scheme. In particular, based on the original optimization problem with a fractional form-objective function, an equivalent optimization problem with a subtractive form-objective function (e.g. $U_R(\textbf{p}, \tilde{\boldsymbol{\alpha}}) - \beta^* U_T(\textbf{p}, \tilde{\boldsymbol{\alpha}})$) can be found such that the same solution can be achieved for both optimization problems. Moreover, \cite{Dinkelbach1967} further implies that the optimal solution is achieved with equality in (\ref{Dinkelbach2}), and thus we could use this equality condition to validate the optimality of the solution. Hence, rather than tackling the original fractional form-objective function (\ref{objectivefa})-(\ref{constraint4fa}), we develop a resource allocation algorithm for the equivalent subtractive form-objective function whilst meeting the conditions in \emph{Proposition 1}. The proposed algorithm is summarized in Table \ref{Dinkelbach method}.

As shown in Table \ref{Dinkelbach method}, the pivotal stage for the proposed Dinkelbach method-based solution is to develop an intermediate resource allocation policy in order to solve the following fixed $\beta$ optimization problem (step 3 in Table \ref{Dinkelbach method}),
\begin{eqnarray}
\label{objectivedin} && \max_{\tilde{\alpha_i},p_i} \sum_{i=1}^{L} \tilde{\alpha}_i \log_2 \left( 1+ \frac{ p_i \lambda_i}{\tilde{\alpha}_i} \right) -  \beta \left( \zeta \sum_{i=1}^{L}p_i + \bar{P}_{sta} + P_{ant}N - \sum_{i=1}^{L} \eta (1-\tilde{\alpha}_i) p_i \lambda_i \right)\\
\label{constraintdin1} && \textrm{s.t.} ~~ \sum_{i=1}^{L} \tilde{\alpha}_i \log_2 \left( 1+ \frac{ p_i \lambda_i}{\tilde{\alpha}_i} \right) \geq R_{min},\\
\label{constraintdin2} && ~~~~~ \sum_{i=1}^{L} \eta (1-\tilde{\alpha}_i) p_i \lambda_i ) \geq E_{min}, \\
\label{constraintdin3} &&~~~~~ \sum_{i=1}^{L}p_i \leq P_{{max}},\\
\label{constraintdin4} &&~~~~~~ p_i \geq 0, \tilde{\alpha}_i \in [0,1], \forall i \in \mathcal{L}.
\end{eqnarray}

\textbf{\emph{Proposition 2:}} \emph{For a given parameter $\beta$, the objective function (\ref{objectivedin}) is strictly and jointly concave in $\tilde{\alpha}_i$ and $p_i$.}\\
\emph{Proof:} See Appendix A.

Therefore, since the objective function is a concave function and the constraint set is also convex, the modified optimization problem in (\ref{objectivedin})-(\ref{constraintdin4}) is in the standard form of a convex programming problem that can be solved by standard numerical methods such as the interior-point method \cite{Boyd}. Hence, problem (\ref{objectivefa})-(\ref{constraint4fa}) can be successfully solved by the proposed convex programming based Dinkelbach method.

\begin{table}\centering
\renewcommand{\arraystretch}{1}  
\begin{tabular} {|l|}
\hline
1)~ Initialize $\beta = 0$, and $\delta$ as the stopping criterion;\\
2)~ \textbf{REPEAT}\\
3)~~~ For a given $\beta$, solve problem (\ref{objectivedin})-(\ref{constraintdin4}) to obtain the\\~~~~~ eigen-channel  assignment and power allocation $\{{\textbf{p}}, {\tilde{\boldsymbol{\alpha}}}\}$;\\
4) ~~~~~~\textbf{IF} $U_R({\textbf{p}}, {\tilde{\boldsymbol{\alpha}}}) - \beta U_T({\textbf{p}}, {\tilde{\boldsymbol{\alpha}}}) \leq \delta$\\
5)~~~~~~~~~~ Convergence $=$ \textbf{TRUE};\\
6)~~~~~~~~~~~\textbf{RETURN} $\{{\textbf{p}^*}, {\tilde{\boldsymbol{\alpha}}}^*\}$ = $\{{\textbf{p}}, {\tilde{\boldsymbol{\alpha}}}\}$ and $\beta^* = \frac{U_R({\textbf{p}}, {\tilde{\boldsymbol{\alpha}}})}{U_T({\textbf{p}}, {\tilde{\boldsymbol{\alpha}}})}$;\\
7) ~~~~~~\textbf{ELSE}\\
8)~~~~~~~~~ Set $\beta = \frac{U_R({\textbf{p}}, {\tilde{\boldsymbol{\alpha}}})}{U_T({\textbf{p}}, {\tilde{\boldsymbol{\alpha}}})}$ and $n = n+1$, Convergence = \textbf{FALSE};\\
9) ~~~~~~\textbf{END IF} \\
10) \textbf{UNTIL} Convergence = \textbf{TRUE}. \\ \hline
\end{tabular} 
\vspace*{0.2em}
\caption{Proposed iterative resource allocation algorithm based on Dinkelbach method}
\label{Dinkelbach method}
\end{table}

\subsection{Joint Eigen-Channel Assignment and Power Allocation (JEAPA)}

The convex programming approach in (\ref{objectivedin})-(\ref{constraintdin4}) is numerically stable, however, its computational complexity depends on the number of optimizing variables, which can be problematic if the number of antenna pair (and hence eigen-channels) is large. In particular, for the method proposed in the last section, we need to update $\beta$ through the Dinkelbach method and apply convex programming-based solution for each iteration to find the optimal eigen-channel assignment and power allocation, and thus the complexity of this scheme is comparably high. Hence, motivated by our previous work in \cite{TangTVTee} where the subcarrier assignment and power allocation process are separated in an OFDMA network, we propose a novel iterative resource allocation approach.

The eigen-channel assignment and power allocation in the modified optimization problem (\ref{objectivefa})-(\ref{constraint4fa}) can hence be separated as follows
\begin{equation}\label{iter_algo}
{\underbrace{\tilde{\boldsymbol{\alpha}}[0] \rightarrow \textbf{p}[0]}_{\textmd{Initialization}}\rightarrow \cdots \underbrace{\tilde{\boldsymbol{\alpha}}[t] \rightarrow \textbf{p}[t]}_{\textmd{Iteration t}}\rightarrow \underbrace{\tilde{\boldsymbol{\alpha}}^{opt} \rightarrow \textbf{p}^{opt}}_{\textmd{Optimal Solution}}.}
\end{equation}
where the number inside the square bracket denotes the iteration number. We proceed by evaluating a feasible solution $(\tilde{\boldsymbol{\alpha}},\textbf{p}[0])$. At the initial moment of each iteration $t$, based on a given power allocation $\textbf{p}[t-1]$ from the last iteration, we solve the eigen-channel assignment problem and obtain the optimal $\tilde{\boldsymbol{\alpha}}[t]$. We then find the optimal power allocation $\textbf{p}[t]$ based on this $\tilde{\boldsymbol{\alpha}}[t]$ obtained from the previous step. This process is repeated until convergence, i.e., no further improvement can be made. Therefore, this iterative resource allocation approach separates the original EE problem under fix $\beta$ into two sub-problems, namely the combinatorial eigen-channel assignment process and the power allocation process. More importantly, the number of variables is decreased by nearly half in each sub-problem, and hence more tractable algorithms could be used to solve the problem.

\subsubsection{Power Allocation under Fixed Eigen-Channel Assignment}

With a fixed eigen-channel assignment $\tilde{\boldsymbol{\alpha}}[t-1]$ obtained from the last iteration, we here attempt to solve the power allocation problem and obtain the optimal allocated power $\textbf{p}[t]$ at iteration $t$. Therefore, the problem in (\ref{objectivefa})-(\ref{constraint4fa}) can now be converted to
\begin{eqnarray}
\label{objectivedineip} && \max_{p_i > 0} \frac{\sum_{i=1}^{L} \tilde{\alpha}_i \log_2 \left( 1+ {p_i \hat{\lambda}_i} \right)}{\zeta \sum_{i=1}^{L}p_i + \bar{P}_{sta} + P_{ant}N - \sum_{i=1}^{L} \eta \check{\lambda}_i p_i }\\
\label{constraintdin1eip} && \textrm{s.t.} ~~ \sum_{i=1}^{L} \tilde{\alpha}_i \log_2 \left( 1+  p_i \hat{\lambda}_i \right) \geq R_{min},\\
\label{constraintdin2eip} && ~~~~~ \sum_{i=1}^{L} \eta  \check{\lambda}_i p_i  \geq E_{min}, \\
\label{constraintdin3eip} &&~~~~~ \sum_{i=1}^{L}p_i \leq P_{{max}}.
\end{eqnarray}
where $\hat{\lambda}_i = \frac{\lambda_i}{\tilde{\alpha}_i}$ and $\check{\lambda}_i = (1-\tilde{\alpha}_i)\lambda_i$ respectively denote the effective channel for data transmission and EH. The above optimization can be solved based on the following proposition.

\textbf{\emph{Proposition 3:}} \emph{With power allocation $p_i, i = 1, 2,\cdots, L$, that satisfies the constraints in (\ref{constraintdin1eip})-(\ref{constraintdin3eip}), the maximum EE, $\psi^*_{EE} = \smash{\max_{p_i > 0}} ~\psi_{EE}(p_i)$, is strictly quasi-concave in $p_i$.} \\
\emph{Proof:} See Appendix B.

The corresponding Lagrangian function can be formulated as
\begin{equation}\label{Lagrangian function1}
{\nonumber G(p_i, \varrho, \kappa, \xi)= \frac{\sum_{i=1}^{L} \tilde{\alpha}_i \log_2 \left( 1+ {p_i \hat{\lambda}_i} \right)}{\zeta \sum_{i=1}^{L}p_i + \bar{P}_{sta} + P_{ant}N - \sum_{i=1}^{L} \eta \check{\lambda}_i p_i }}
\end{equation}
\begin{equation}\label{Lagrangian function_cont1}
{ + \varrho \left( \sum_{i=1}^{L} \tilde{\alpha}_i \log_2 \left( 1 +  p_i \hat{\lambda}_i  \right) - R_{min} \right)+ \kappa \left( \sum_{i=1}^{L} \eta  \check{\lambda}_i p_i - E_{min} \right)+ \xi \left( P_{{max}}-\sum_{i=1}^{L}p_i \right) }
\end{equation}
where $\varrho \geq 0$, $\kappa \geq 0$ and $\xi \geq 0$ are the Lagrangian multipliers associated with the minimum rate, minimum harvested energy, and maximum transmit power constraints, respectively. Thus, the dual objective function is written as
\begin{equation}\label{gp}
    {l(\varrho, \kappa, \xi)=\max_{p_i}~G(p_i, \varrho, \kappa, \xi)}.
\end{equation}
The dual problem is accordingly given by
\begin{equation}\label{dual_gp}
{\min_{\varrho, \kappa, \xi}~ l(\varrho, \kappa, \xi) ~~~s.t. ~~~\varrho \geq 0, \kappa \geq 0, \xi \geq 0.}
\end{equation}
By invoking the Karush-Kuhn-Tucker (KKT) conditions, the optimal solution set $\{p_1, \cdots, p_L\}$  can be obtained through the gradient ascent
algorithm in \cite{Zhang09}. In each iteration, step $p_i$ can be updated sequentially according to its gradient direction of the Lagrangian function (\ref{Lagrangian function_cont1}) as follows
\begin{equation}\label{gradient_n}
    {\nonumber \nabla_{p_i}G:= \frac{\tilde{\alpha}_i \hat{\lambda}_i \log_2e }{(1+p_i\check{\lambda}_i )[\sum_{i=1}^{L} (\zeta-\eta\check{\lambda}_i )p_i +P_{fix} ]} - \frac{(\zeta-\eta\check{\lambda}_i)\sum_{i=1}^{L} \tilde{\alpha}_i \log_2 (1+ {p_i \hat{\lambda}_i} )}{[\sum_{i=1}^{L} (\zeta-\eta\check{\lambda}_i )p_i +P_{fix} ]^2}  }
\end{equation}
\begin{equation}\label{gradienttt}
    {+  \frac{\varrho\tilde{\alpha}_i \hat{\lambda}_i \log_2e}{1+p_i\check{\lambda}_i } + \kappa \eta \check{\lambda}_i - \xi}
\end{equation}
\begin{equation}\label{updatett}
    {p_i(n) = [p_i(n-1) + \varepsilon(n-1) \nabla_{p_i}G]^+},
\end{equation}
where $P_{fix} = \bar{P}_{sta} + P_{ant}N$, $\varepsilon$ represents the step size of iteration $n(n \in\{1, 2, \cdots , I_{max}\})$, with $I_{max}$ being the maximum number of iterations. The step size update should meet the condition
\begin{equation}\label{step4}
    {\sum_{n=1}^{\infty}\varepsilon(n) =  \infty,~\lim_{n\rightarrow \infty} \varepsilon = 0.}
\end{equation}

Once the optimal $p_i^*$ is obtained using (\ref{gradienttt}) and (\ref{updatett}), we can determine the optimal dual variables $\varrho, \kappa, \xi$. Since the Lagrangian function $l(\varrho, \kappa, \xi)$ is convex over $\varrho, \kappa, \xi$, a one-dimensional searching approach can be applied here. However, the gradient approach is not always available given that $l(\varrho, \kappa, \xi)$ is not guaranteed to be differentiable. On the other hand, we can use the well-known sub-gradient approach in order to update the dual variables $\varrho, \kappa, \xi$. In particular, the sub-gradient direction is described as in the following lemma.

\textbf{\emph{Lemma 1.}} \emph{$   \sum_{i=1}^{L} \tilde{\alpha}_i \log_2 (1+  p_i \hat{\lambda}_i  ) - R_{min}$, $ \sum_{i=1}^{L} \eta  p_i \check{\lambda}_i - E_{min}$ and $P_{max}- \sum_{i=1}^{L}p_i$ are the subgradient of the dual objective function $l(\varrho, \kappa, \xi)$, respectively.}\\
\emph{Proof:} Please refer to \cite{Zhang09} for a proof of \emph{Lemma 1}.

The dual variables can therefore be updated using
\begin{equation}\label{update2p}
    {\varrho(n) = \left [\varrho(n-1) + \omega(n-1) \left( R_{min}-  \sum_{i=1}^{L} \tilde{\alpha}_i \log_2 (1+  p_i \hat{\lambda}_i  )   \right) \right]^+},
\end{equation}
\begin{equation}\label{update3p}
    {\kappa(n) = \left[ \kappa(n-1) + \omega(n-1) \left( E_{min}- \sum_{i=1}^{L} \eta  p_i \check{\lambda}_i \right) \right]^+},
\end{equation}
\begin{equation}\label{update4p}
    {\xi(n) = \left [\xi(n-1) + \omega(n-1) \left( \sum_{i=1}^{L}p_i - P_{max} \right) \right]^+},
\end{equation}
where $\omega$ is used to denote the step size which satisfies the condition in (\ref{step4}).

\subsubsection{Eigen-Channel Assignment under Fixed Power Allocation}

Next, we consider a fixed power allocation $\textbf{p}[t-1]$ obtained from the last iteration, and attempt to solve the eigen-channel assignment problem to obtain $\tilde{\boldsymbol{\alpha}}[t]$. Therefore, the problem in (\ref{objectivedin})-(\ref{constraintdin4}) is converted to
\begin{eqnarray}
\label{objectivedinei} && \max_{\tilde{\alpha}_i} \frac{\sum_{i=1}^{L} \tilde{\alpha}_i \log_2 \left( 1+ \frac{\tilde{\lambda}_i}{\tilde{\alpha}_i} \right)}{\zeta \sum_{i=1}^{L}p_i + \bar{P}_{sta} + P_{ant}N - \sum_{i=1}^{L} \eta (1-\tilde{\alpha}_i) \tilde{\lambda}_i}\\
\label{constraintdin1ei} && \textrm{s.t.} ~~ \sum_{i=1}^{L} \tilde{\alpha}_i \log_2 \left( 1+ \frac{ \tilde{\lambda}_i}{\tilde{\alpha}_i} \right) \geq R_{min},\\
\label{constraintdin2ei} && ~~~~~ \sum_{i=1}^{L} \eta (1-\tilde{\alpha}_i) \tilde{\lambda}_i  \geq E_{min}, \\
\label{constraintdin3ei} &&~~~~~~ 0 \leq \tilde{\alpha}_i \leq 1, \forall i \in \mathcal{L},
\end{eqnarray}
where $\tilde{\lambda}_i = p_i \lambda_i$ denotes the effective channel (including the power allocated). Similar to the case of power allocation under fixed eigen-channel assignment, the above optimization problem is quasi-concave with respect to $\tilde{\alpha}_i$. The corresponding Lagrangian function can therefore be expressed as
\begin{equation}\label{Lagrangian function}
{\nonumber L(\tilde{\alpha}_i, \nu_i, \tau, \varsigma)= \frac{\sum_{i=1}^{L} \tilde{\alpha}_i \log_2 \left( 1+ \frac{\tilde{\lambda}_i}{\tilde{\alpha}_i} \right)}{\zeta \sum_{i=1}^{L}p_i + \bar{P}_{sta} + P_{ant}N - \sum_{i=1}^{L} \eta (1-\tilde{\alpha}_i) \tilde{\lambda}_i}}
\end{equation}
\begin{equation}\label{Lagrangian function_cont}
{ + \sum_{i=1}^{L} \nu_i(1-\tilde{\alpha}_i) + \tau \left( \sum_{i=1}^{L} \tilde{\alpha}_i \log_2 \left( 1 + \frac{ \tilde{\lambda}_i}{\tilde{\alpha}_i} \right)-R_{min} \right) + \varsigma \left( \sum_{i=1}^{L} \eta ( 1-\tilde{\alpha}_i) \tilde{\lambda}_i - E_{min} \right) }
\end{equation}
where $\nu_i \geq 0$, $\tau \geq 0$ and $\varsigma \geq 0$ are the Lagrangian multipliers associated with the constraints in terms of eigen-channel assignment, minimum rate, and minimum harvested energy, respectively. Thus, the dual objective function and the dual problem can be respectively written as
\begin{equation}\label{g}
    {g(\nu_i, \tau, \varsigma)=\max_{\tilde{\alpha}_i}~L(\tilde{\alpha}_i, \nu_i, \tau, \varsigma)}
\end{equation}
and
\begin{equation}\label{dual_g}
{\min_{\nu_i, \tau, \varsigma} g(\nu_i, \tau, \varsigma) ~~~s.t. ~~~\tau \geq 0, \varsigma \geq 0, \nu_i \geq 0, i \in \mathcal{L}.}
\end{equation}
The optimal solutions of the sub-problems, $\tilde{\alpha}_i^*$, can be obtained through the gradient of the Lagrangian function (\ref{Lagrangian function_cont}) with respect to $\tilde{\alpha}_i$ under KKT conditions as 
\begin{equation}\label{gradientdd}
    {\nonumber \nabla_{\tilde{\alpha}_i}L:= \left[ \log_2 \left( 1+\frac{\tilde{\lambda}_i}{\tilde{\alpha}_i} \right) -  \frac{ \tilde{\lambda}_i \log_2e}{\tilde{\alpha}_i+\tilde{\lambda}_i} \right] \left[ \left( \sum_{i=1}^{L} \eta \tilde{\lambda}_i \tilde{\alpha}_i + \tilde{P}_{fix} \right)^{-1}+\tau \right] }
\end{equation}
\begin{equation}\label{gradient}
    {-\eta \tilde{\lambda}_i \left( \sum_{i=1}^{L} \eta \tilde{\lambda}_i \tilde{\alpha}_i + \tilde{P}_{fix} \right)^{-2} \sum_{i=1}^{L}\log_2 \left( 1+\frac{\tilde{\lambda}_i}{\tilde{\alpha}_i} \right) - \nu_i-\varsigma \eta \tilde{\lambda}_i}
\end{equation}
\begin{equation}\label{update}
    {\tilde{\alpha}_i(n) = [\tilde{\alpha}_i(n-1) + \epsilon(n-1) \nabla_{\tilde{\alpha}_i}L]^+},
\end{equation}
where $\tilde{P}_{fix} = \sum_{i=1}^{L}p_i + \bar{P}_{sta} + P_{ant}N - \sum_{i=1}^{L}\eta\tilde{\lambda}_i$, $\epsilon$ represents the step size of iteration, and satisfy a similar condition as in (\ref{step4}). The dual variables $\nu_i, \tau, \varsigma$ are then updated using
\begin{equation}\label{update2}
    {\tau(n) = \left[ \tau(n-1) + \phi(n-1)(R_{min}-\sum_{i=1}^{L} \tilde{\alpha}_i \log_2 \left(1+ \frac{ \tilde{\lambda}_i}{\tilde{\alpha}_i} ) \right) \right]^+},
\end{equation}
\begin{equation}\label{update3}
    {\varsigma(n) = \left[ \varsigma(n-1) + \phi(n-1) \left( E_{min}-\sum_{i=1}^{L} \eta (1-\tilde{\alpha}_i) \tilde{\lambda}_i \right) \right]^+},
\end{equation}
\begin{equation}\label{update4}
    {\nu_i(n) = [\nu_i(n-1) + \phi(n-1)(\tilde{\alpha}_i - 1)]^+, \forall i \in \mathcal{L}},
\end{equation}
where $\phi$ is the step size which satisfy the condition in (\ref{step4}).

\subsection{Multi-Objective Optimization Low-Complexity Approach (MOO-LC)}

The iterative solution proposed in the previous subsection is more efficient compared to the convex programming-based Dinkelbach method; however, it still requires a large number of iterations if there exists many antennas (eigen-channels). With the goal of reducing the resource-intensity, we propose a low-complexity heuristic algorithm based on the idea of MOO to avoid approximating the eigen-channel assignment and power allocation in an iterative fashion. In particular, we first determine the ``appropriate" power allocation assuming that each eigen-channel is allocated for ID and EH at the same time. Then, based on the power allocation result, we apply the eigen-channel assignment scheme proposed in the previous subsection to determine the optimal eigen-channel assignment. Finally, with the allocated eigen-channels for ID and EH, the proposed power allocation strategy is applied again to further improve the EE performance.

It is intuitive to consider uniform power allocation for the initial phase of the power allocation strategy. However, it is observed that one constraint usually dominates over the other, acquiring the “best” eigen-channels, i.e., those with the largest eigenvalues. Hence, a ``fair''  uniform power allocation approach may not be suitable for the proposed SS-based MIMO SWIPT system. Based on the MOO approach in \cite{Coello}, we propose a compromised power allocation algorithm considering both data transmission and EH. In general, a constrained MOO problem is defined as follows \cite{Coello}
\begin{eqnarray}
\label{objectivemo} && \min_{\textbf{x}}~~ F(\textbf{x}) = (f_1(\textbf{x}), f_2(\textbf{x}),\cdots, f_k(\textbf{x}))\\
\label{constraint1mo} && \textrm{s.t.} ~~~~ g_i(\textbf{x}) \leq 0,  i = 1,2,\cdots,m,\\
\label{constraint2mo} &&~~~~~~~~ h_j(\textbf{x}) = 0, j = 1,2,\cdots,n,
\end{eqnarray}
\noindent where $F(\textbf{x})$, $g_i(\textbf{x})$ and $h_j(\textbf{x})$ respectively denote the set of objective functions, the set of inequality constraints and the set of equality constraints. Given that the objective functions are usually conflicting, a constrained MOO problem is able to simultaneously optimize $k$ objective functions. Moreover, we can apply the weighted-sum (scalarization) method to solve the MOO problem where the multiple objectives are combined and transformed into a single-objective scalar function. Specifically, the weighted-sum method optimizes a positively weighted convex sum of the objectives, that is
\begin{eqnarray}
\label{objectivec} && \min_{\gamma_l, \textbf{x}}~~ \sum_{l = 1}^{k} \gamma_l f_l(\textbf{x})\\
\label{constraint1c} && \textrm{s.t.} ~~~~ \sum_{l = 1}^{k}\gamma_l = 1, ~ \gamma_l > 0,~ l = 1,2,\cdots,k,\\
\label{constraint2c} &&~~~~~~~~ g_i(\textbf{x}) \leq 0,  i = 1,2,\cdots,m.
\end{eqnarray}
This represents a new optimization problem with a unique objective function (in weighted-sum-form). It can be proved that the minimizer of this weighted-sum single-objective function is an efficient solution for the original MOO problem \cite{Coello}, i.e., its image belongs to the Pareto curve.

In this work, since both sum-rate and power transfer constraints are taken into account, we consider both of them as our objective functions.
In addition, we need to unify the two objectives since energy and rate cannot be compared directly. Similar to \cite{Qisun2014}, a statistical value is employed here to model the potential transmission rate provided by the harvested power
\begin{equation}\label{cstore}
    {C_{EH} =  \theta \sum_{i=1}^{L} \eta  p_i \lambda_i }.
\end{equation}
where $\theta$ represents the efficiency of transferring the harvested energy to data transmission. Hence, we have the following MOO problem in order to determine the power initialization
\begin{eqnarray}
\label{objectivecsw} && \max_{p_i>0} ~\left\{\sum_{i=1}^{L}  \log_2 (1+  p_i \lambda_i ),\theta \sum_{i=1}^{L} \eta p_i \lambda_i \right\}\\
\label{constraint1csw} && \textrm{s.t.} ~~~~ \sum_{i=1}^{L}p_i \leq P_{{max}}.
\end{eqnarray}
The above constrained MOO problem can be transformed into a single-objective function, written as
\begin{eqnarray}
\label{objectivecs} && \max_{p_i>0} ~\sum_{i=1}^{L} \gamma_1 \log_2 (1+ { p_i \lambda_i} )+\gamma_2\theta \sum_{i=1}^{L} \eta p_i \lambda_i \\
\label{constraint2cs} && \textrm{s.t.} ~~~\sum_{i=1}^{L}p_i \leq P_{{max}}
\end{eqnarray}
where $\gamma_1 = \frac{R_{min}}{R_{min} + \theta E_{min}}$ and $\gamma_2 = \frac{\theta E_{min}}{R_{min} + \theta E_{min}}$. Clearly, the objective function (\ref{objectivecs}) is a linear combination of concave and affine functions with respect to $p_i$; hence it is concave. The problem in (\ref{objectivecs})-(\ref{constraint2cs}) is in turn convex. As a result, according to the KKT conditions, the optimal solutions can be obtained using
\begin{equation}\label{water_filling_sub}
    { p_i^* = \left[ \frac{\log_2e}{ (R_{min}+\theta E_{min})/(\varphi R_{min} ) - \theta^2 \eta \lambda_i E_{min} /R_{min} } - \frac{1}{\lambda_i} \right]^+ }
\end{equation}
where $\varphi \geq 0$ is the Lagrangian multiplier associated with the maximum transmit power constraint. A bisection approach \cite{Tang2013} can be employed here to update $\varphi$ where the sub-gradient is $P_{max}- \sum_{i=1}^{L}p_i$.

With this power allocation strategy, we can then apply the eigen-channel assignment scheme proposed in the previous subsection to determine the optimal eigen-channel allocation for either ID or EH. Finally, we need to perform the proposed power allocation algorithm in the previous subsection to further improve the EE.

\section{Active Receive Antenna Selection}

In this section, we further study the receive antenna selection approach to explore the achievable EE in a SS-based SWIPT MIMO system. Activating all the receive antennas is always optimal in terms of throughput optimization, but not for EE optimization. This is because although activating more receive antennas will achieve a higher sum-rate as well as harvesting more energy, it comes at a cost of higher circuit power consumption (dynamic parts which based on the configurations of the active links). Therefore, there exists a trade-off between the power consumption cost and the sum-rate and harvested energy gain. As a result, receive antenna selection is essential in terms of maximizing EE.

For the SS-based MIMO SWIPT system, it is intuitive to conclude that the optimal receive antenna selection strategy is the exhaustive search. Specifically, for each possible receive antenna set $\chi \in \{1,2,\cdots,N_R\}$, we obtain the EE based on the proposed joint eigen-channel assignment and power allocation algorithm in section III, and then select the optimal active receive antenna set as
\begin{equation}\label{as}
    { \chi^{opt} = \arg \max_{\chi \in \{1,2,\cdots,N_R\}} \psi_{EE}(\chi)  }.
\end{equation}
However, the computational complexity of this exhaustive search scheme is too high to implement in practice. Therefore, developing low-complexity approaches with low-complexity is necessary per discussed in the following.

Since the optimal solution has to calculate the EE for all possible antenna sets, this implies that the antenna selection process is separated from the eigen-channel assignment and power allocation procedures. In other words, the receive antenna selection process is not connected with SS (eigen-channel) for ID and EH. Therefore, with a given number of receive antenna $N = |\chi|$, our aim is to select the set of receive antennas that maximizes the EE of the system which is used for either ID or EH. We therefore arrive at the following.

\textbf{\emph{Proposition 4.}} \emph{With a fixed transmit power $P$, the maximum EE of the system where ID and EH are operated at the same time, can be achieved using the following MOO problem
\begin{eqnarray}
\label{objectiveAS} && \max_{\chi:|\chi|=N,~ \textbf{Q}_{\chi}> 0} ~\left\{  \frac{\log\det(\textbf{I}_{N} + \textbf{H}_\chi \textbf{Q}_\chi \textbf{H}_\chi^H)}{\zeta \textmd{tr}(\textbf{Q}_\chi) +   \bar{P}_{sta} + P_{ant}N   } ,     \frac{\theta \eta \textmd{tr}(\textbf{H}_\chi \textbf{Q}_\chi \textbf{H}_\chi^H)}{\zeta \textmd{tr}(\textbf{Q}_\chi) +   \bar{P}_{sta} + P_{ant}N}      \right\}\\
\label{constraint1AS} && \textmd{s.t.} ~~~~ \textmd{tr}(\textbf{Q}_\chi) = P.
\end{eqnarray}
where $\frac{\log\det(\textbf{I}_{N} + \textbf{H}_\chi \textbf{Q}_\chi \textbf{H}_\chi^H)}{\zeta \textmd{tr}(\textbf{Q}_\chi) +   \bar{P}_{sta} + P_{ant}N   }$ represents the EE of the conventional ID MIMO system whilst $\frac{\theta \eta \textmd{tr}(\textbf{H}_\chi \textbf{Q}_\chi \textbf{H}_\chi^H)}{\zeta \textmd{tr}(\textbf{Q}_\chi) +   \bar{P}_{sta} + P_{ant}N} $ denotes the EE of the EH MIMO system.} \\
\emph{Proof:} See Appendix C.

Considering equal transmit power allocation at each antenna, we can transform (\ref{objectiveAS}) using
\begin{equation} \label{objectiveAS1}
{\nonumber \max_{\chi:|\chi|=N,~ P> 0} ~\left\{  \frac{\log\det(\textbf{I}_{N} + \frac{P}{N} \textbf{H}_\chi \textbf{H}_\chi^H)}{\zeta P +   P_{sta} + P_{ant}N   } ,     \frac{\theta \eta \frac{P}{N} \textmd{tr}(\textbf{H}_\chi \textbf{H}_\chi^H)}{\zeta P +   \bar{P}_{sta} + P_{ant}N} \right\} }
\end{equation}
\begin{equation} \label{objectiveAS2}
{= \max_{\chi:|\chi|=N} ~\{  \det( \textbf{H}_\chi \textbf{H}_\chi^H),     \textmd{tr}(\textbf{H}_\chi \textbf{H}_\chi^H)      \}. }
\end{equation}
However, calculating the channel matrix determinant or the trace of the channel matrix requires a large number of computations, especially when the system is equipped with a large number of antennas. Therefore, instead of directly applying determinant or trace operations to the channel matrix, we here incorporate the Frobenius-norm of the channel matrix in order to reduce the computational complexity. Although the Frobenius-norm of the channel cannot directly characterize the capacity and harvested energy precisely, it is related to the throughput and harvested energy by demonstrating the overall energy of the channel \cite{ZukangShen}. As a result, the selection criterion for the SS-based MIMO SWIPT system is based on
\begin{equation}\label{app_eEE5}
{ \textmd{sort}_{1 \leq n \leq N}  ~||\textbf{h}_{n}||_{F}^{2}}
\end{equation}
where $\textbf{h}_{n}$ denotes the $n$-th row of the channel matrix ${\textbf{H}}$, which represents the channel quality of the $n$-th receive antenna. After sorting, the receive antenna set is selected from the first $N_R$ rows of the sorted matrix. We then only need to perform the proposed eigen-channel assignment and power allocation algorithm in Section III to maximize EE. This process is repeated until all the receive antenna number has been investigated. The complete solution to the EE optimization problem for MIMO SWIPT system with SS technique is summarized in Table \ref{The complete solution to the EE optimization problem}.

\begin{table}\centering
\renewcommand{\arraystretch}{1}  
\begin{tabular} {|l|}
\hline
1)~~Initialization: sort the antennas using (\ref{app_eEE5});\\
2)~~\textbf{For} $N=1:N_R$\\
3)~~~~Find the best $N_R$ receive antennas based on Frobenius norm method;\\
4)~~~~Calculate the optimal EE using the proposed eigen-channel assignment \\~~~~~ and power allocation algorithms in Section III, denoted as $\psi_{EE}^{opt}({N})$; \\
5)~~\textbf{End For} \\
6)~~Compare all the EE in the buffer and select the set of receive antenna \\~~~ that maximizes the EE. \\ \hline
\end{tabular} 
\vspace*{0.2em}
\caption{The complete solution to the EE optimization problem with antenna selection} 
\label{The complete solution to the EE optimization problem}
\end{table}

For the exhaustive search approach, the search size is $\sum_{i=1}^{N_R} \frac{N_R!}{i!(N_R-i)!}$. Therefore, the joint eigen-channel assignment and power allocation algorithm run-time increases exponentially as a function of the receive antenna number $N_R$. On the other hand, the proposed norm-based selection approach requires the joint eigen-channel assignment and power allocation algorithm to run $N_R$ times. Hence, is significantly less resource-intensive and thus more suitable for implementation in practice.

\section{Simulation Results}

\begin{figure}\centering
\includegraphics{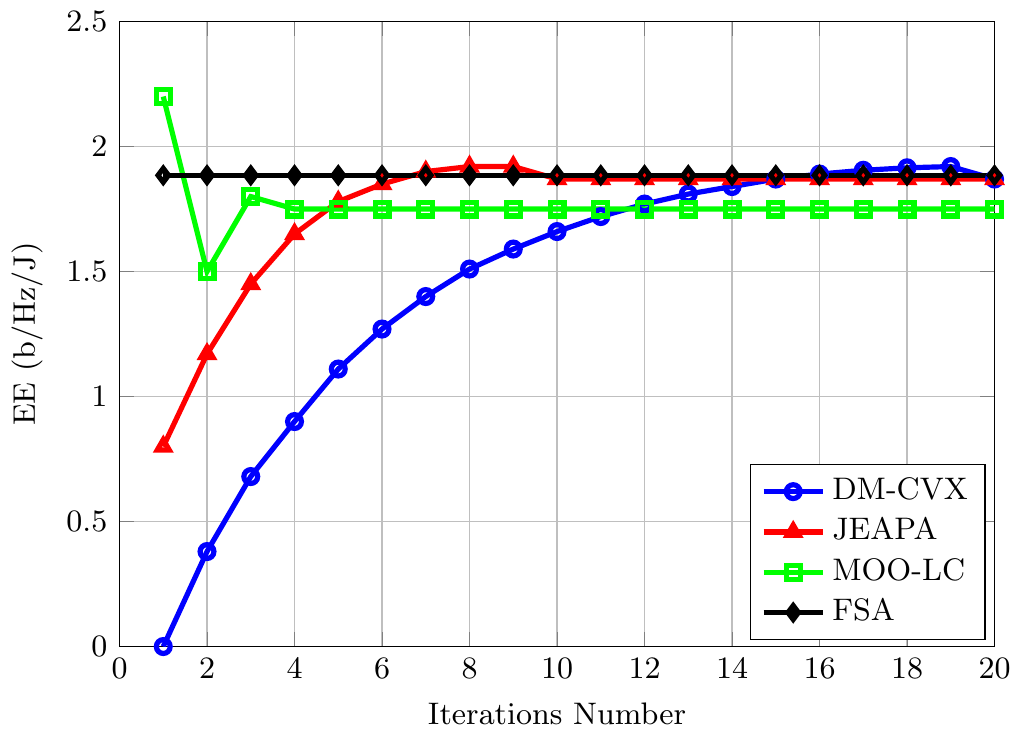}
 \caption{Convergence behavior of the proposed eigen-channel assignment and power allocation approaches (fixed antenna set).}
 \label{new_conv} 
\end{figure}

\begin{figure}\centering
 \includegraphics{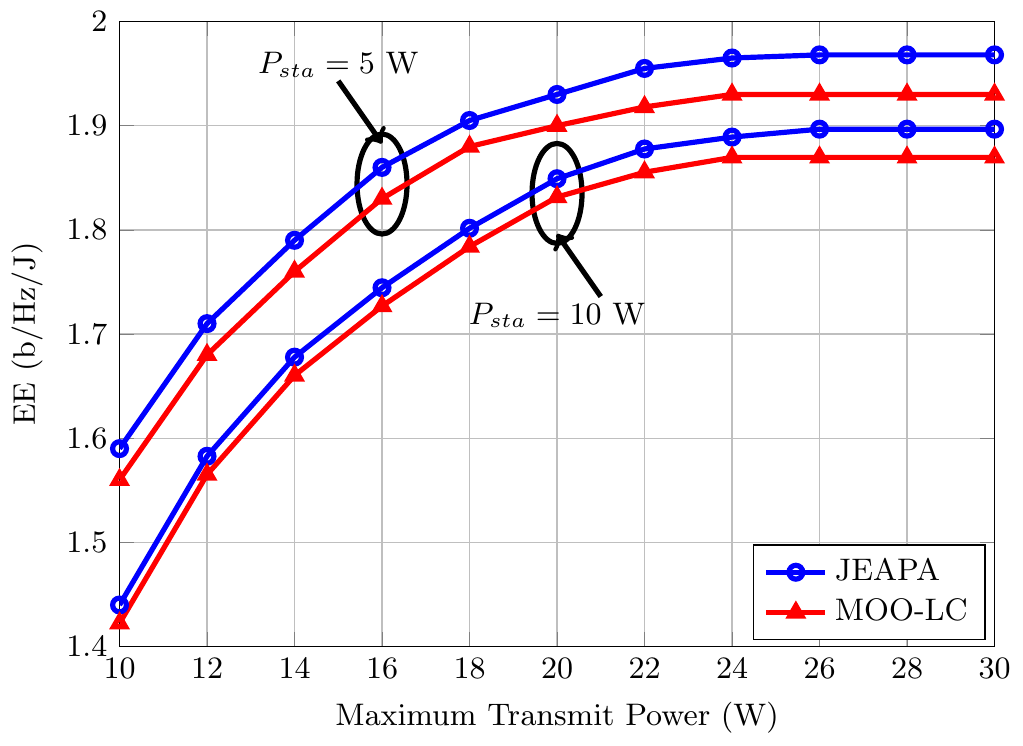}
 \caption{The performance of the proposed Dinkelbach method-based joint eigen-channel assignment and power allocation algorithm with different static circuit power (EE vs transmit power constraint).}
 \label{transmit_power} 
\end{figure}

In this section, we present numerical results to verify the theoretical findings and analyze the effectiveness of the proposed approaches. In our simulations, the total number of transmit and receive antennas are respectively $N_T = 8$ and $N_R = 8$. In addition, the drain efficiency of the power amplifier $\zeta$ is set to $38\%$ whereas the EH efficiency is taken to be $\eta = 10\%$. The static circuit power at the transmitter $P_{sta}$ is assumed to be 5 W and the dynamic power consumption proportional to the number of antennas $P_{ant}$ is set to be 1 W. It should be noted that these system parameters are merely chosen to demonstrate the EE performance in an example and can easily be modified to any other values depending on the specific scenario.

In the first simulation, the convergence behavior of the proposed joint eigen-channel assignment and power allocation algorithms are studied. We fix the number of active receive antennas to $N = N_R = 8$. For convenience, we denote the full-search-based approach as FSA. The convergence behavior of these inner-layer solution is evaluated by illustrating how the EE performance behaves with the number of iterations. As shown in Fig. \ref{new_conv}, JEAPA converges to a stable value which is achieved by DM-CVX, but with a faster convergence speed (almost reduce by half). More importantly, the EE achieved by the proposed DM-CVX and JEAPA are very close to the FSA. This demonstrates that the proposed algorithms can efficiently approach the optimal EE. Moreover, it is observed that there is a drop on EE at the 10-th and 20-th iteration for JEAPA and DM-CVX respectively. This is because after these solutions converge, the possibly fractional $\tilde{\alpha}_i^*$ is rounded to either 0 or 1 and the proposed power allocation algorithm has been performed again to get the maximum EE for the round-off case. This result further coincides with our theoretical findings where both schemes are upper-bound solutions due to the relaxation of $\alpha_i$. Finally, the proposed MOO-LC converges to a lower but acceptable EE compared to the optimal FSA within four iterations, and hence this scheme is suitable to implement in practice.

\begin{figure}\centering
 \includegraphics{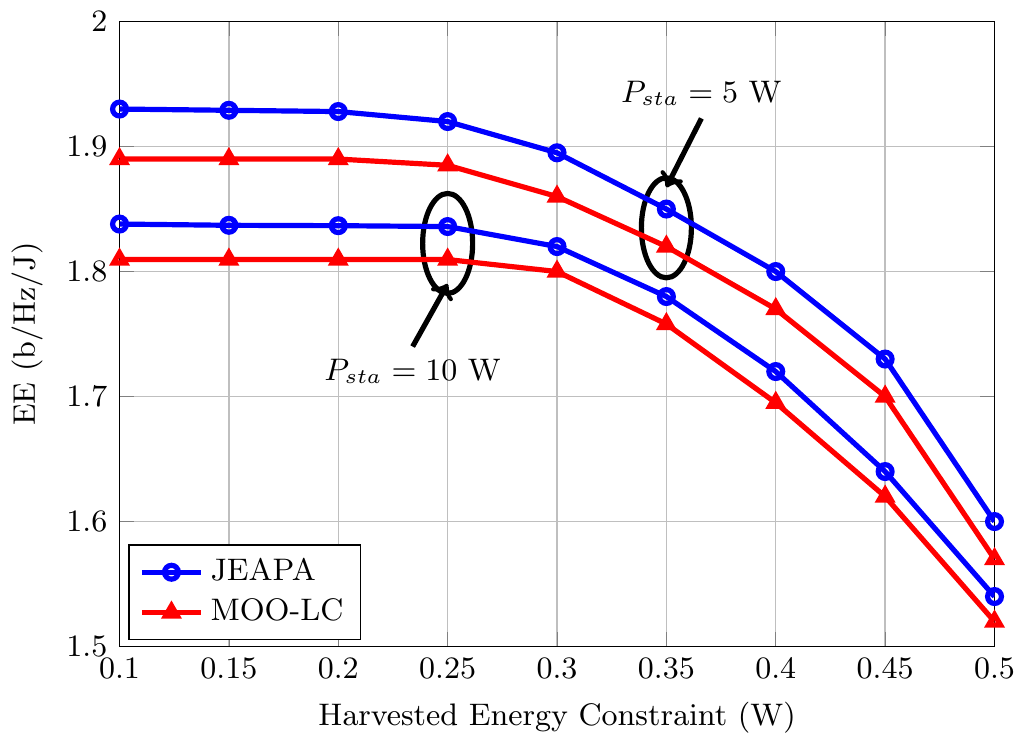}
 \caption{The performance of the proposed Dinkelbach method-based joint eigen-channel assignment and power allocation algorithm with different static circuit power (EE vs EH constraint).}
 \label{MIN_EH} 
\end{figure}

In the next simulation, JEAPA and MOO-LC under different constraints are evaluated and presented in Fig. \ref{transmit_power}, Fig. \ref{rate} and Fig. \ref{MIN_EH}. The proposed algorithms under different maximum transmit power constraints are evaluated first. As can be seen from Fig. \ref{transmit_power}, the EE achieved by our proposed JEAPA algorithm is monotonically non-decreasing with respect to the maximum transmit power constraint $P_{max}$. Particularly, the EE increases in the lower transmit power constraint region, i.e., $10 < P_{max} < 20$ W, and then saturates when $P_{max} > 20$ W due to the fact that a balance between the system EE and the total power consumption can be achieved. Fig. \ref{transmit_power} also compares and indicates the influence of static circuit power on the EE-transmission power relation. From there, as expected, EE decreases with increased circuit power due to the higher power consumption. We next show in Fig. \ref{rate} the maximum EE under different minimum rate requirement and different circuit power settings. Note that the optimal EE is the same up to a certain minimum rate requirement, but drops afterwards. This is because when the minimum rate requirement is low, the required transmit power is also low. Thus, the most energy efficient design is to operate at a higher transmit power in order to achieve the optimal EE. We investigate the EE versus the minimum required harvested energy for the proposed EE maximization algorithm. As shown in Fig. \ref{MIN_EH}, similar trend is observed to the case of increasing the minimum rate demand. In particular, the optimal EE is the same up to a certain minimum required harvested energy, but drops afterwards.

\begin{figure}\centering
 \includegraphics{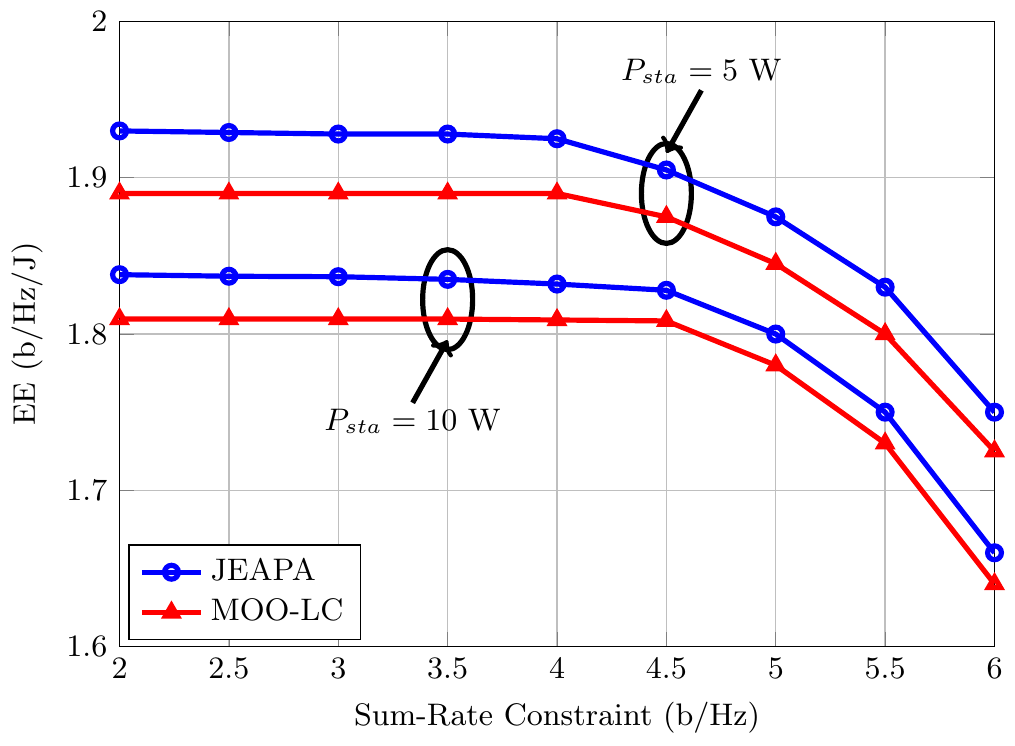}
 \caption{The performance of the proposed Dinkelbach method-based joint eigen-channel assignment and power allocation algorithm with different static circuit power (EE vs rate constraint).}
 \label{rate} 
\end{figure}

\begin{figure}\centering
 \includegraphics{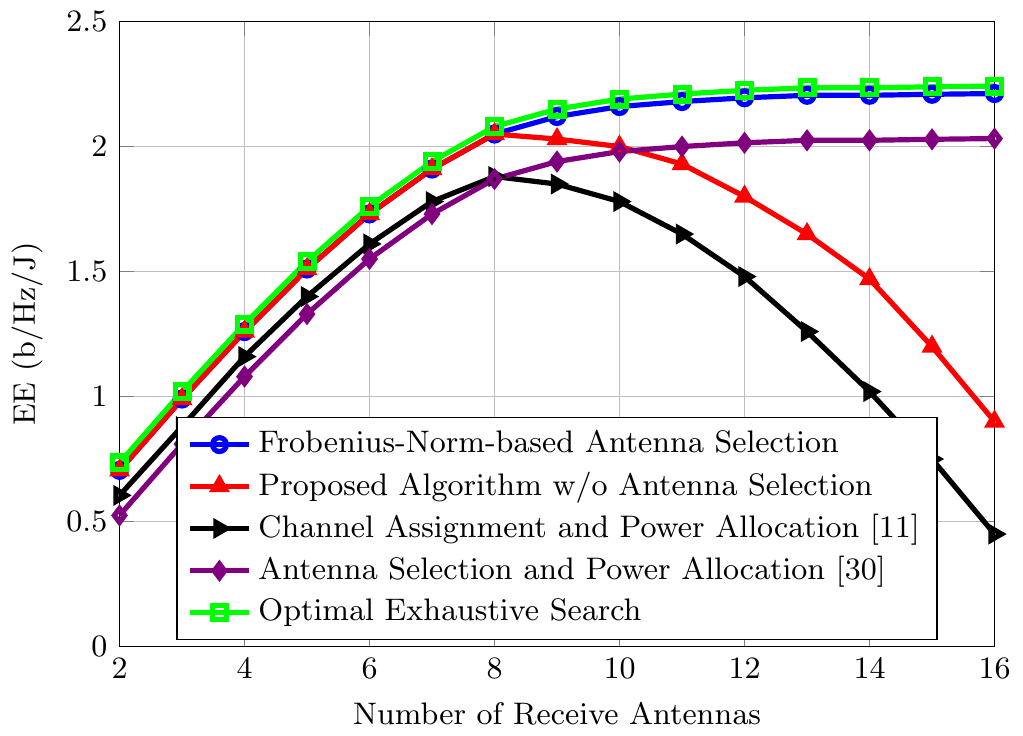}
 \caption{Energy efficiency versus the maximum transmit power allowance for the proposed extended BC-MAC duality-based EE maximization algorithm with different minimum required power transfer.}
 \label{all_scheme} 
\end{figure}

Finally, we evaluate the performance of the proposed norm-based receive antenna selection algorithm for the SS-based MIMO SWIPT system. To show the EE gain, we compare with the scheme that maximizes the EE but without EH \cite{TangTVT}, and the scheme in \cite{Timotheou2015} which minimizes the transmit power in SS-based MIMO SWIPT system without considering antenna selection strategy. We also show the performance of our proposed scheme but without antenna selection. In other words, this scheme always employs all the receive antennas. As shown in Fig. \ref{all_scheme}, the EE achieved by our proposed norm-based antenna selection approach is monotonically non-decreasing with respect to the number of active receive antenna $N$. Particularly, the EE increases linearly with an increasing $N$ in the lower region, i.e., $2 < N < 8$, and then saturates when $N > 10$ as a balance between the system EE and the spatial (eigen-channel) gain is achieved. Furthermore, the EE achieved by the proposed norm-based selection approach outperform the EE achieved in \cite{TangTVT} and \cite{Timotheou2015}, and is very close to the optimal exhaustive search approach; but with a lower complexity. Interestingly, for the case of higher total number of active receive antennas, i.e., $N \geq 10$, the EE achieved by the proposed algorithm without considering antenna selection is lower than that of the scheme proposed in \cite{TangTVT}. This implies that the EE gain achieved by EH cannot compensate the cost of activating redundant antennas.

\section{Conclusions}
In this paper, we address the EE optimization problem for MIMO SWIPT system with SS-based receiver. Considering a practical linear power model where the number of active receive antennas, transmit power, and power transfer are taken into consideration, our target is to maximize the EE whilst satisfying certain constraints in terms minimum sum-rate and power transfer. The EE optimization problem, which involves joint optimization of the eigen-channel assignment, power allocation, and active receive antenna set, is non-convex, and thus extremely difficult to tackle directly. Hence, to obtain a feasible solution for this problem, we propose to separate the antenna selection procedure with the eigen-channel assignment and power allocation operation.
In particular, under fixed receive antenna set, near-optimal convex programming-based Dinkelbach method, iterative joint eigen-channel assignment and power allocation algorithm, and MOO-based low-complexity solutions are developed. We then study antenna selection to further explore the achievable EE and accordingly provide optimal exhaustive search and Frobenius-norm-based dynamic selection schemes. Numerical results illustrate that the proposed joint antenna selection and SS-based approach outperforms state-of-the-art schemes in terms of improving the EE performance of the MIMO SWIPT \nolinebreak[4] system.

\section*{Appendix A}
\begin{center}
    \textsc{Proof of Proposition 2}
 \end{center}
To show that the objective function in (\ref{objectivedin}) is concave, we can reformulate it using
\begin{equation}\label{ob_change}
{\nonumber \sum_{i=1}^{L} \tilde{\alpha}_i \log_2 \left(1+ \frac{ p_i \lambda_i}{\tilde{\alpha}_i} \right) -  \beta \left( \zeta \sum_{i=1}^{L}p_i + P_{sta} + P_{ant}N - \sum_{i=1}^{L} \eta (1-\tilde{\alpha}_i) p_i \lambda_i \right)}
\end{equation}
\begin{equation}\label{ob_change_cont}
{\nonumber = \sum_{i=1}^{L} \tilde{\alpha}_i \log_2 \left( 1+ \frac{ p_i \lambda_i}{\tilde{\alpha}_i} \right) + \sum_{i=1}^{L}(\beta\eta\lambda_i - \beta\zeta )p_i - \sum_{i=1}^{L}\beta\eta\lambda_i \tilde{\alpha}_i p_i - \sum_{i=1}^{L} \frac{\beta (P_{sta} + P_{ant}N)}{L}}
\end{equation}
\begin{equation}\label{ob_change_cont1}
{\nonumber = \sum_{i=1}^{L} \left\{\tilde{\alpha}_i \log_2 (1+ \frac{ p_i \lambda_i}{\tilde{\alpha}_i} ) + (\beta\eta\lambda_i - \beta\zeta )p_i - \beta\eta\lambda_i \tilde{\alpha}_i p_i - \frac{\beta (P_{sta} + P_{ant}N)}{L} \right\} }
\end{equation}
\begin{equation}\label{ob_change_cont2}
{= \sum_{i=1}^{L} g(\tilde{\alpha}_i, p_i)}.
\end{equation}
Given that linear combination does not affect the convexity property, we are interested in the behavior of $g(\tilde{\alpha}_i, p_i)$. A two-dimensional function is concave if and only if its restriction to a line is concave \cite{Boyd04}. Defining $x = \tilde{\alpha}_i$ and $y = p_i$, and let $y = ax+b$, $g(\tilde{\alpha}_i, p_i)$ can be reformulated as
\begin{equation}\label{refor}
{\nonumber f(x,y) =   x \log \left( 1+\frac{y}{x} \right)+ y - xy}
\end{equation}
\begin{equation}\label{refor_c1}
{= x \log \left( 1+a+\frac{b}{x} \right)-ax^2+(a-b)x+b}.
\end{equation}
Further denoting the above function as $\bar{f}(x)$, its concavity is verified by taking the second derivative such that
\begin{equation}\label{second_der}
{\nabla^2 \bar{f}(x) =  \frac{b/(1+a)}{x+b/(1+a)} \left( \frac{1}{x+b/(1+a)} - \frac{1}{x} \right) - 2a } .
\end{equation}
Since $\tilde{\alpha}_i$ and $p_i$ are non-negative variables and $p_i$ is non-decreasing with increasing $\tilde{\alpha}_i$, we thus have $a > 0$. Furthermore, as $\frac{b/(1+a)}{x+b/(1+a)} (\frac{1}{x+b/(1+a)} - \frac{1}{x}  )$ is always negative, the second derivative of $\bar{f}(x)$ is always negative. Consequently, $\bar{f}(x)$ is concave when $x \geq 0$, and $f(x,y)$ is concave and hence (\ref{objectivedin}) is strictly and jointly concave in $\tilde{\alpha}_i$ and $p_i$. \hspace{\fill} $\blacksquare$

\section*{Appendix B}
\begin{center}
    \textsc{Proof of Proposition 3}
 \end{center}
To prove ${\psi}^*_{EE}(p_i)$ is a quasi-concave function, we denote the superlevel sets of ${\psi}^*_{EE}(p_i)$ as
\begin{equation}\label{quasiconcave_example}
{\mathcal{S}_\pi = \left\{ \max \{R_{min}^{-1}(p_i), E_{min}^{-1}(p_i)\} \leq  \sum_{i=1}^{L} p_i \leq  P_{{max}} | {\psi}^*_{EE}(p_i) \geq \pi \right\}.}
\end{equation}
For any real number $\pi$, if the convexity for $\mathcal{S}_\pi$ holds, ${\psi}^*_{EE}(p_i)$ is strictly quasi-concave in $p_i$ \cite{Boyd04}. Therefore, we here divide the proof into two cases. For the case of $\pi< 0$, since EE is always positive and hence there are no points on the counter, ${\psi}^*_{EE}(p_i) = \pi$. For the case of $\pi \geq 0$, $\psi_{EE}$ can be rewritten as
\begin{equation}\label{EE_proof}
{\psi_{EE} = \frac{C(p_i)}{\sum_{i=1}^{L} (\zeta-\eta\check{\lambda}_i)p_i + P_{fix}}}.
\end{equation}
Hence, $\mathcal{S}_\pi$ is equivalent to $\pi \sum_{i=1}^{L}(\zeta-\eta\check{\lambda}_i)p_i  + \pi P_{fix} - C(p_i) \leq 0$. Since it has been proven that $C(p_i)$ is concave with respect to $p_i$ in Appendix A, and $\pi (\zeta-\eta\check{\lambda}_i)p_i$ is an affine function with respect to $p_i$, the convexity property for $\mathcal{S}_\pi$ holds and ${\psi}^*_{EE}(p_i)$ is therefore strictly quasi-concave in $p_i$. This completes the proof of \emph{Proposition 3}.\hspace{\fill} $\blacksquare$

\section*{Appendix C}
\begin{center}
    \textsc{Proof of Proposition 4}
 \end{center}
To proceed, we first look at the following EE maximization problem where ID and EH are operating at the same time
\begin{eqnarray}
\label{ob_EE_all} && \max_{\chi:|\chi|=N,~ \textbf{Q}_{\chi}> 0} ~  \frac{\log\det(\textbf{I}_{N} + \textbf{H}_\chi \textbf{Q}_\chi \textbf{H}_\chi^H)}{\zeta \textmd{tr}(\textbf{Q}_\chi) +   P_{sta} + P_{ant}N -  \eta \textmd{tr}(\textbf{H}_\chi \textbf{Q}_\chi \textbf{H}_\chi^H) }\\
\label{constraintob_EE_all} && \textrm{s.t.} ~~~~ \textmd{tr}(\textbf{Q}_\chi) = P.
\end{eqnarray}
By leveraging on the theory of non-linear fractional programming in \cite{Dinkelbach1967}, we can use the Dinkelbach method to solve this non-linear fractional programming problem. Specifically, we transform the fractional form objective function into a numerator-denominator subtractive form. The key step for the Dinkelbach-based method is to solve the optimization problem for a given parameter in each iteration and then update this parameter accordingly. As a result, under a given parameter $\varpi$, the objective function in (\ref{ob_EE_all}) can be reformulated as
\begin{equation}\label{reob_EE_all}
{\max_{\chi:|\chi|=N, \textbf{Q}_{\chi}> 0} ~  \log\det(\textbf{I}_{N} + \textbf{H}_\chi \textbf{Q}_\chi \textbf{H}_\chi^H) + \varpi \eta \textmd{tr}(\textbf{H}_\chi \textbf{Q}_\chi \textbf{H}_\chi^H) - \varpi(\zeta \textmd{tr}(\textbf{Q}_\chi) +   P_{sta} + P_{ant}N).}
\end{equation}
Since the transmit power $P$ is considered to be fixed in this case, $\varpi(\zeta \textmd{tr}(\textbf{Q}_\chi) + P_{sta} + P_{ant}N)$ is a constant. Hence, 
\begin{equation}\label{reob_EE_all_further}
{\max_{\chi:|\chi|=N, \textbf{Q}_{\chi}> 0} ~  \gamma_1 \log\det(\textbf{I}_{N} + \textbf{H}_\chi \textbf{Q}_\chi \textbf{H}_\chi^H) + \gamma_2 \eta \textmd{tr}(\textbf{H}_\chi \textbf{Q}_\chi \textbf{H}_\chi^H)}
\end{equation}
where $\gamma_1 = 1$ and $\gamma_2 = \varpi$. This weighted-sum optimization problem forms an efficient solution for the original MOO problem in (\ref{objectiveAS})-(\ref{constraint1AS}). This completes the proof. \hspace{\fill} $\blacksquare$

\bibliographystyle{IEEEtran}
\bibliography{IEEEabrv,references}

\begin{thebibliography}{10}
\providecommand{\url}[1]{#1}
\csname url@samestyle\endcsname
\providecommand{\newblock}{\relax}
\providecommand{\bibinfo}[2]{#2}
\providecommand{\BIBentrySTDinterwordspacing}{\spaceskip=0pt\relax}
\providecommand{\BIBentryALTinterwordstretchfactor}{4}
\providecommand{\BIBentryALTinterwordspacing}{\spaceskip=\fontdimen2\font plus
\BIBentryALTinterwordstretchfactor\fontdimen3\font minus
  \fontdimen4\font\relax}
\providecommand{\BIBforeignlanguage}[2]{{%
\expandafter\ifx\csname l@#1\endcsname\relax
\typeout{** WARNING: IEEEtran.bst: No hyphenation pattern has been}%
\typeout{** loaded for the language `#1'. Using the pattern for}%
\typeout{** the default language instead.}%
\else
\language=\csname l@#1\endcsname
\fi
#2}}
\providecommand{\BIBdecl}{\relax}
\BIBdecl

\bibitem{6623062}
X.~Zhou, R.~Zhang, and C.~K. Ho, ``Wireless information and power transfer:
  Architecture design and rate-energy tradeoff,'' \emph{IEEE Trans. Commun.},
  vol.~61, no.~11, pp. 4754--4767, Nov. 2013.

\bibitem{6951347}
X.~Lu, P.~Wang, D.~Niyato, D.~I. Kim, and Z.~Han, ``Wireless networks with {RF}
  energy harvesting: A contemporary survey,'' \emph{IEEE Commun. Surveys
  Tuts.}, vol.~17, no.~2, pp. 757--789, Second Quart. 2015.

\bibitem{6845056}
E.~Hossain, M.~Rasti, H.~Tabassum, and A.~Abdelnasser, ``Evolution toward {5G}
  multi-tier cellular wireless networks: An interference management
  perspective,'' \emph{IEEE Trans. Wireless Commun.}, vol.~21, no.~3, pp.
  118--127, June 2014.

\bibitem{4595260}
L.~R. Varshney, ``Transporting information and energy simultaneously,'' in
  \emph{Proc. 2008 IEEE Int. Symp. Inf. Theory}, July 2008, pp. 1612--1616.

\bibitem{6489506}
R.~Zhang and C.~K. Ho, ``{MIMO} broadcasting for simultaneous wireless
  information and power transfer,'' \emph{IEEE Trans. Wireless Commun.},
  vol.~12, no.~5, pp. 1989--2001, May 2013.

\bibitem{Xiangz2012}
Z.~Xiang and M.~Tao, ``Robust beamforming for wireless information and power
  transmission,'' \emph{IEEE Wireless Commun. Lett.}, vol.~1, no.~2, pp.
  372--375, Aug. 2012.

\bibitem{Liu2013}
L.~Liu, R.~Zhang, , and K.-C. Chua, ``Wireless information transfer with
  opportunistic energy harvesting,'' \emph{IEEE Trans. Wireless Commun.},
  vol.~12, no.~2, pp. 288--300, Jan.~2013.

\bibitem{AANasir}
A.~A. Nasir, X.~Zhou, S.~Durrani, and R.~A. Kennedy, ``Relaying protocols for
  wireless energy harvesting and information processing,'' \emph{IEEE Trans.
  Wireless Commun.}, vol.~12, no.~7, pp. 3622--3636, July 2013.

\bibitem{Liu2013c}
L.~Liu, R.~Zhang, , and K.~C. Chua, ``Wireless information and power transfer:
  a dynamic power splitting approach,'' \emph{IEEE Trans. Commun.}, vol.~61,
  no.~9, pp. 3990--4001, Sep.~2013.

\bibitem{ZhouX14}
X.~Zhou, R.~Zhang, and C.~K. Ho, ``Wireless information and power transfer in
  multiuser {OFDM} systems,'' \emph{IEEE Trans. Wireless Commun.}, vol.~13,
  no.~4, pp. 2282--2294, Apr. 2014.

\bibitem{Timotheou2015}
S.~Timotheou, I.~Krikidis, S.~Karachontzitisand, and K.~Berberidis, ``Spatial
  domain simultaneous information and power transfer for {MIMO} channels,''
  \emph{IEEE Trans. Wireless Commun.}, vol.~14, no.~8, pp. 4115--4128,
  Aug.~2015.

\bibitem{TangJSAC}
J.~Tang, D.~K.~C. So, E.~Alsusa, K.~A. Hamdi, and A.~Shojaeifard, ``Resource
  allocation for energy efficiency optimization in heterogeneous networks,''
  \emph{IEEE Journal Sel. Areas in Commun.}, vol.~33, no.~10, pp. 2104--2117,
  Oct.~2015.

\bibitem{Hasan2011}
Z.~Hasan, H.~Boostanimehr, and V.~K. Bhargava, ``Green cellular networks: A
  survey, some research issues and challenges,'' \emph{IEEE Commun. Surveys
  Tutorials}, vol.~13, no.~4, pp. 524--540, Fourth Quart. 2011.

\bibitem{Chen2011}
Y.~Chen, S.~Zhang, S.~Xu, and G.~Y. Li, ``Fundamental tradeoffs on green
  wireless networks,'' \emph{IEEE Commun. Mag.}, vol.~49, no.~6, pp. 30--37,
  June~2011.

\bibitem{Miao2010}
G.~Miao, N.~Himayat, and G.~Y. Li, ``Energy-efficient link adaptation in
  frequency-selective channels,'' \emph{IEEE Trans. Commun.}, vol.~58, no.~2,
  pp. 545--554, Feb.~2010.

\bibitem{TangRE}
J.~Tang, D.~K.~C. So, E.~Alsusa, and K.~A. Hamdi, ``Resource efficiency: A new
  paradigm on energy efficiency and spectral efficiency tradeoff,'' \emph{IEEE
  Trans. Wireless Commun.}, vol.~13, no.~8, pp. 4656--4669, Aug.~2014.

\bibitem{6918448}
A.~Shojaeifard, K.~A. Hamdi, E.~Alsusa, D.~K.~C. So, and J.~Tang, ``A unified
  model for the design and analysis of spatially-correlated load-aware
  {HetNets},'' \emph{IEEE Trans. Commun.}, vol.~62, no.~11, pp. 1--16, Nov.
  2014.

\bibitem{TangTCOM}
J.~Tang, D.~K.~C. So, E.~Alsusa, K.~A. Hamdi, and A.~Shojaeifard, ``Energy
  efficiency optimization with interference alignment in multi-cell {MIMO}
  interfering broadcast channels,'' \emph{IEEE Trans. Commun.}, vol.~63, no.~7,
  pp. 2486--2499, July 2015.

\bibitem{armanEE}
A.~Shojaeifard, K.-K. Wong, K.~A. Hamdi, E.~Alsusa, D.~K.~C. So, and J.~Tang,
  ``Stochastic geometric analysis of energy-efficient dense cellular
  networks,'' \emph{arXiv:1610.06846}, 2016.

\bibitem{QShi}
Q.~Shi, C.~Peng, W.~Xu, and M.~Hong, ``Energy efficiency optimization for
  \uppercase{MISO} \uppercase{SWIPT} systems with zero-forcing beamforming,''
  \emph{IEEE Trans. Sig. Process.}, vol.~64, no.~4, pp. 842--854, Feb. 2016.

\bibitem{HeS}
S.~He, Y.~Huang, W.~Chen, S.~Jin, H.~Wang, and L.~Yang, ``Energy efficient
  coordinated precoding design for a multicell system with \uppercase{RF}
  energy harvesting,'' \emph{EURASIP J. Wireless Commun. Netw.}, no.~1, pp.
  1--12, Mar. 2015.

\bibitem{6774838}
M.~R.~A. Khandaker and K.~K. Wong, ``{SWIPT} in {MISO} multicasting systems,''
  \emph{IEEE Wireless Commun. Lett.}, vol.~3, no.~3, pp. 277--280, June 2014.

\bibitem{7322196}
J.~Liao, M.~R.~A. Khandaker, and K.~K. Wong, ``Robust power-splitting {SWIPT}
  beamforming for broadcast channels,'' \emph{IEEE Commun. Lett.}, vol.~20,
  no.~1, pp. 181--184, Jan. 2016.

\bibitem{TangsubmitTVT}
J.~Tang, D.~K.~C. So, Y.~Fu, Z.~Zhou, B.~Li, Y.~Xiang, and Z.~Wu, ``Energy
  efficient resource allocation for \uppercase{MIMO} broadcast channels with
  simultaneous wireless information and power transfer,'' \emph{IEEE Trans.
  Veh. Tech.}, submitted 2016.

\bibitem{Xiong2011}
C.~Xiong, G.~Li, S.~Zhang, Y.~Chen, and S.~Xu, ``Energy- and
  spectral-efficiency tradeoff in downlink \uppercase{OFDMA} networks,''
  \emph{IEEE Trans. Wireless Commun.}, vol.~10, no.~11, pp. 3874--3886,
  Nov.~2011.

\bibitem{7478073}
A.~Shojaeifard, K.~A. Hamdi, E.~Alsusa, D.~K.~C. So, J.~Tang, and K.~K. Wong,
  ``Design, modeling, and performance analysis of multi-antenna heterogeneous
  cellular networks,'' \emph{IEEE Trans. Commun.}, vol.~64, no.~7, pp.
  3104--3118, July 2016.

\bibitem{Goldsmith2005}
A.~Goldsmith, \emph{Wireless Communications}.\hskip 1em plus 0.5em minus
  0.4em\relax Stanford University, 2005.

\bibitem{7277029}
A.~Shojaeifard, K.~A. Hamdi, E.~Alsusa, D.~K.~C. So, and J.~Tang, ``Exact
  {SINR} statistics in the presence of heterogeneous interferers,'' \emph{IEEE
  Trans. Inf. Theory}, vol.~61, no.~12, pp. 6759--6773, Dec. 2015.

\bibitem{6781609}
D.~W.~K. Ng, E.~S. Lo, and R.~Schober, ``Robust beamforming for secure
  communication in systems with wireless information and power transfer,''
  \emph{IEEE Trans. Wireless Commun.}, vol.~13, no.~8, pp. 4599--4615, Aug.
  2014.

\bibitem{TangTVT}
J.~Tang, D.~K.~C. So, E.~Alsusa, K.~A. Hamdi, and A.~Shojaeifard, ``On the
  energy efficiency-spectral efficiency trade-off in \uppercase{MIMO-OFDMA}
  broadcast channels,'' \emph{IEEE Trans. Veh. Tech.}, vol.~65, no.~7, pp.
  5185--5199, July~2016.

\bibitem{Boyd04}
S.~Boyd and L.~Vandenberghe, \emph{Convex Optimization.}\hskip 1em plus 0.5em
  minus 0.4em\relax Cambridge University Press, Cambridge, UK, 2004.

\bibitem{Yu2002}
W.~Yu and J.~M. Coffi, ``\uppercase{FMDA} capacity of \uppercase{G}aussian
  multiple-access channel with \uppercase{ISI},'' \emph{IEEE Trans on Commun.},
  vol.~50, no.~1, pp. 102--111, Jan. 2002.

\bibitem{Dinkelbach1967}
W.~Dinkelbach, ``On nonlinear fractional programming,'' \emph{Management
  Science}, vol.~13, pp. 492--498, Mar.~1967.

\bibitem{Boyd}
S.~Boyd, \emph{Branch and Bound Methods.}\hskip 1em plus 0.5em minus
  0.4em\relax Stanford University.

\bibitem{TangTVTee}
J.~Tang, D.~K.~C. So, E.~Alsusa, K.~A. Hamdi, A.~Shojaeifard, and K.-K. Wong,
  ``Energy-efficient heterogeneous cellular networks with spectrum underlay and
  overlay access,'' \emph{arXiv:1610.09683}, 2016.

\bibitem{Zhang09}
L.~Zhang, Y.~Xin, and Y.~C. Liang, ``Weighted sum rate optimization for
  cognitive radio \uppercase{MIMO} broadcast channels,'' \emph{IEEE Trans.
  Wireless Commun.}, vol.~8, no.~9, pp. 2950--2959, Jun.~2009.

\bibitem{Coello}
G.~B.~L. C.~A.~C.~Coello and D.~A.~V. Veldhuizen, \emph{Evolutionary Algorithms
  for Solving Multi-\uppercase{O}bjective Problems.}\hskip 1em plus 0.5em minus
  0.4em\relax New \uppercase{Y}ork: \uppercase{S}pringer, 2007.

\bibitem{Qisun2014}
Q.~Sun, L.~Li, and J.~Mao, ``Simultaneous infomation and power transfer scheme
  for energy efficient {MIMO} systems,'' \emph{IEEE Commun. Lett.}, vol.~18,
  no.~4, pp. 600--603, Apr.~2014.

\bibitem{Tang2013}
J.~Tang and S.~Lambotharan, ``Interference alignment techniques for
  \uppercase{MIMO} multi-cell interfering broadcast channels,'' \emph{IEEE
  Trans. Commun.}, vol.~61, no.~1, pp. 164--175, Jan.~2013.

\bibitem{ZukangShen}
Z.~Shen, R.~Chen, J.~G. Andrews, R.~W. Heath, and B.~L. Evans, ``Low complexity
  user selection algorithms for multiuser \uppercase{MIMO} systems with block
  diagonalization,'' \emph{IEEE Trans. Sig. Lett.}, vol.~54, no.~9, pp.
  3658--3663, Sep.~2006.

\end{thebibliography}

\end{document}